\documentclass{amsart}
\usepackage{dsfont}
\usepackage{pstricks,pst-node,pst-text,pst-3d, pstricks-add}
\usepackage{subfig}
\usepackage{enumitem}
\usepackage{layout}

\author{Florian Simatos}
\email{florian.simatos@inria.fr}
\urladdr{http://www-rocq.inria.fr/$\sim$simatos}
\address{INRIA, domaine de Voluceau, B.P.~105, 78153 Le Chesnay Cedex, France}

\title[A variant of the Recoil Growth algorithm]
{\mbox{A variant of the Recoil Growth}\\\mbox{algorithm to generate multi-polymer systems}}

\date{\today}

\def\acc{\mathrm{acc}}
\def\C{\mathrm{c}}
\def\new{\mathrm{new}}
\def\old{\mathrm{old}}

\def\n{\mathrm{n}}
\def\o{\mathrm{o}}
\def\RG{\mathrm{g}}
\def\U{\mathrm{u}}

\def\G{\mathcal{G}}

\def\R{RG$^*$ }
\def\Z{\mathds{Z}}

\newtheorem{definition}{Definition}

\newtheorem{theorem}{Theorem}
\newtheorem{proposition}{Proposition}
\newtheorem{conjecture}{Conjecture}
\newtheorem{lemma}{Lemma}

\setenumerate[1]{label=\arabic*.}
\setenumerate[3]{label=\roman*.}

\begin{document}

\bibliographystyle{amsplain}

\begin{abstract}
The Recoil Growth algorithm, proposed in 1999 by Consta \emph{et al.}, is one of the most efficient algorithm available in the literature to sample from a multi-polymer system. Such problems are closely related to the generation of self-avoiding paths. In this paper, we study a variant of the original Recoil Growth algorithm, where we constrain the generation of a new polymer to take place on a specific class of graphs. This makes it possible to make a fine trade-off between computational cost and success rate. We moreover give a simple proof for a lower bound on the irreducibility of this new algorithm, which applies to the original algorithm as well.
\end{abstract}

\maketitle

\hrule

\vspace{-0.2cm}

\tableofcontents

\vspace{-8mm}

\hrule

\bigskip

\section{Introduction}

Designing an algorithm that efficiently samples from a multi-polymer system according to a given probability distribution is the focus of much research activity in chemical physics~\cite{Continuous, CBMC-1}. The state of a multi-polymer system being a collection of self-avoiding paths that do not overlap each other, this problem is closely related to the classical problem in computer science of generating self-avoiding paths. The state space of such systems is huge, especially in high dimension or if the paths are long, which makes these problems hard. The main issue consists of defining an algorithm that both keeps the computational cost low and converges rapidly to the sampling distribution. For multi-polymer systems, various approaches have been suggested to tackle this problem, among which is the Recoil Growth (RG) algorithm, meant to be one of the most efficient algorithm currently available in the literature. For a precise description of this algorithm, the reader  is referred to~\cite{Consta}\footnote{The present paper is self-contained, and does not require prior knowledge on the RG algorithm.}. In the present paper (which is an extended version of~\cite{simatos}), we define and analyze a variant of the RG algorithm, to which we refer as the \R algorithm. The main ideas of both the RG and the \R algorithms are to be found in two important classes of algorithms, which we briefly describe below: the Metropolis algorithm~\cite{metropolis-orig} and the auxiliary variable method~\cite{aux-1, aux-3}.

\subsection*{The Metropolis algorithm}

The Metropolis algorithm is a very generic algorithm that approximately samples according to a probability distribution $\pi$ defined on some finite state space $\mathcal X$. It takes as other input a Markov Chain $P$, and constructs a reversible Markov Chain with $\pi$ as stationary distribution. In the long-run, we can therefore sample from a distribution that is arbitrarily close to $\pi$. The implicit idea is that it is hard to sample from $\pi$, whereas it is easy to generate $P$, for instance by local modifications of a state. The new Markov Chain is built thanks to the following rejection procedure, in which we use the notation $A(x,y) = \pi(x) P(x,y)$. Starting from $x \in \mathcal X$, choose as candidate for the next state some $y \in \mathcal X$ with probability $P(x, y)$. If $A(y,x) \geq A(x,y)$, then go to $y$. Otherwise, flip a coin with bias $A(y,x) / A(x,y)$: if tails, go to $y$, and if heads, stay in $x$. By doing so, the probability of going from $x$ to $y \neq x$ is $P(x,y)\cdot\min\big(1, A(y,x)/A(x,y)\big)$, and it is easy to see that this Markov chain has $\pi$ as reversible distribution. An important remark is that for the rejection procedure, only the ratio $A(y,x)/A(x,y)$ matters. In particular, it is sufficient to know $\pi$ up to its normalizing constant, what is very useful in many concrete applications. For instance, one could wish to sample complex combinatorial objects uniformly: in this case, one does not need to know the cardinality of these objects.

Under a more global look, the Markov Chain $P$ can be seen as an exogenous process that, at each step of the algorithm, proposes a candidate for the next step. In general, the stationary distribution of $P$ is not $\pi$, and so the role of the rejection procedure aims at compensating for this bias: by suitably rejecting or accepting this candidate, the new Markov Chain can be shown to be reversible with $\pi$ as stationary distribution. Reversibility is an important feature here, because it makes it very easy to show that $\pi$ is indeed the stationary distribution. Otherwise, since the objects under consideration are usually fairly complex, proving stationarity could be very challenging.

\subsection*{The auxiliary variables method}

The auxiliary variable method is a conceptually easy extension of the Metropolis algorithm: instead of sampling from $\mathcal X$ according to $\pi$, it is sometimes easier to sample from a pair $\mathcal X \times \mathcal Y$ according to some distribution $\widetilde \pi(\cdot, \cdot)$ which has $\pi$ as marginal. In such cases, the idea is to apply the Metropolis algorithm to $\widetilde \pi$, and then to recover $\pi$ by summation. The new variable $y \in \mathcal Y$ is referred to as the ``auxiliary variable''. As we will see, the concept of \emph{underlying graph} in the \R algorithm is very close to an auxiliary variable; underlying graphs are central in the \R algorithm, and are introduced not because they make the sampling easier or more natural, but because they reduce the computational cost.

\subsection*{The \R algorithm}

\begin{figure}
\begin{center}
  \psset{unit=0.06,linewidth=0.6pt,linecolor=black}
  \begin{pspicture}(137,137)
\psline(77,3)(77,4)(77,5)(78,5)(79,5)(80,5)(80,6)(80,7)(80,8)(79,8)(79,7)(78,7)(78,6)(77,6)(77,7)(77,8)(76,8)(76,7)(76,6)(76,5)(76,4)(75,4)(74,4)(74,5)
\psline(76,57)(76,58)(77,58)(77,59)(78,59)(79,59)(79,60)(78,60)(78,61)(78,62)(77,62)(77,61)(77,60)(76,60)(75,60)(75,61)(76,61)(76,62)(76,63)(77,63)(78,63)(78,64)(79,64)(79,65)
\psline(94,16)(95,16)(95,17)(96,17)(97,17)(97,16)(97,15)(97,14)(96,14)(95,14)(95,13)(94,13)(94,14)(94,15)(93,15)(93,14)(93,13)(92,13)(91,13)(90,13)(89,13)(88,13)(88,12)(89,12)
\psline(18,65)(18,66)(17,66)(17,67)(18,67)(18,68)(18,69)(17,69)(17,70)(16,70)(16,69)(15,69)(15,68)(14,68)(13,68)(13,67)(13,66)(13,65)(13,64)(12,64)(12,63)(12,62)(12,61)(12,60)
\psline(112,114)(113,114)(113,115)(114,115)(114,114)(114,113)(114,112)(114,111)(115,111)(115,110)(116,110)(116,111)(116,112)(116,113)(116,114)(115,114)(115,115)(115,116)(114,116)(114,117)(113,117)(112,117)(112,118)(111,118)
\psline(74,129)(73,129)(72,129)(72,128)(73,128)(74,128)(74,127)(74,126)(73,126)(72,126)(72,125)(72,124)(72,123)(73,123)(73,124)(74,124)(75,124)(75,125)(75,126)(76,126)(76,127)(76,128)(76,129)(77,129)
\psline(119,90)(118,90)(118,89)(117,89)(117,90)(116,90)(116,89)(116,88)(116,87)(117,87)(117,86)(118,86)(118,87)(118,88)(119,88)(119,89)(120,89)(121,89)(122,89)(122,88)(122,87)(122,86)(122,85)(123,85)
\psline(56,124)(56,123)(56,122)(56,121)(55,121)(55,122)(54,122)(54,121)(53,121)(53,120)(54,120)(55,120)(55,119)(54,119)(54,118)(54,117)(54,116)(53,116)(52,116)(51,116)(51,117)(52,117)(53,117)(53,118)
\psline(55,91)(54,91)(54,92)(55,92)(56,92)(57,92)(57,91)(57,90)(56,90)(55,90)(55,89)(56,89)(56,88)(57,88)(58,88)(58,89)(58,90)(59,90)(60,90)(61,90)(61,89)(60,89)(60,88)(60,87)
\psline(47,118)(48,118)(48,117)(48,116)(49,116)(49,115)(49,114)(48,114)(48,113)(47,113)(46,113)(46,114)(46,115)(45,115)(45,116)(45,117)(46,117)(46,118)(46,119)(47,119)(47,120)(46,120)(45,120)(45,121)
\psline(52,98)(52,99)(52,100)(52,101)(51,101)(50,101)(50,100)(51,100)(51,99)(51,98)(50,98)(50,97)(49,97)(48,97)(48,98)(48,99)(49,99)(49,100)(49,101)(48,101)(48,102)(49,102)(49,103)(48,103)
\psline(113,73)(112,73)(111,73)(111,72)(112,72)(113,72)(114,72)(114,73)(115,73)(115,72)(115,71)(115,70)(114,70)(114,71)(113,71)(113,70)(113,69)(113,68)(112,68)(112,67)(111,67)(110,67)(110,66)(109,66)
\psline(26,67)(26,68)(26,69)(26,70)(27,70)(27,69)(27,68)(28,68)(28,67)(29,67)(29,66)(29,65)(29,64)(29,63)(30,63)(30,62)(29,62)(29,61)(29,60)(30,60)(30,59)(31,59)(31,60)(32,60)
\psline(113,106)(112,106)(111,106)(111,105)(112,105)(113,105)(114,105)(114,106)(114,107)(113,107)(112,107)(111,107)(110,107)(110,108)(109,108)(109,109)(110,109)(110,110)(109,110)(109,111)(108,111)(107,111)(107,112)(106,112)
\psline(84,96)(84,95)(84,94)(83,94)(83,95)(82,95)(82,96)(82,97)(82,98)(82,99)(81,99)(80,99)(79,99)(79,98)(79,97)(78,97)(78,96)(79,96)(79,95)(80,95)(80,94)(80,93)(81,93)(81,94)
\psline(104,78)(104,77)(103,77)(103,76)(102,76)(101,76)(101,75)(101,74)(101,73)(101,72)(102,72)(103,72)(104,72)(104,71)(103,71)(102,71)(102,70)(102,69)(101,69)(100,69)(100,68)(100,67)(100,66)(101,66)
\psline(25,75)(25,76)(25,77)(24,77)(24,78)(23,78)(22,78)(22,77)(22,76)(21,76)(21,77)(21,78)(20,78)(19,78)(19,79)(19,80)(19,81)(20,81)(21,81)(22,81)(22,82)(23,82)(23,83)(24,83)
\psline(21,5)(22,5)(22,6)(23,6)(23,5)(24,5)(24,6)(24,7)(24,8)(25,8)(26,8)(26,9)(27,9)(27,10)(28,10)(28,9)(28,8)(29,8)(30,8)(30,7)(29,7)(29,6)(29,5)(30,5)
\psline(14,82)(15,82)(16,82)(16,83)(16,84)(16,85)(17,85)(17,84)(18,84)(18,83)(19,83)(19,84)(19,85)(20,85)(21,85)(22,85)(22,84)(21,84)(20,84)(20,83)(21,83)(21,82)(20,82)(19,82)
\psline(12,41)(12,42)(13,42)(14,42)(15,42)(15,41)(14,41)(14,40)(13,40)(12,40)(12,39)(12,38)(13,38)(14,38)(14,37)(14,36)(14,35)(14,34)(14,33)(13,33)(13,34)(13,35)(13,36)(12,36)
\psline(86,128)(85,128)(85,129)(84,129)(84,130)(85,130)(85,131)(85,132)(86,132)(86,133)(85,133)(85,134)(84,134)(83,134)(83,135)(83,136)(84,136)(85,136)(86,136)
\psline(88,136)(89,136)
\psline(131,26)(131,25)(130,25)(130,24)(131,24)(131,23)(131,22)(131,21)(132,21)(133,21)(134,21)(134,20)(135,20)
\psline(0,20)(1,20)(1,21)(0,21)
\psline(136,22)(135,22)(134,22)(133,22)(133,23)
\psline(56,77)(56,78)(56,79)(55,79)(54,79)(53,79)(52,79)(52,78)(51,78)(51,77)(50,77)(50,76)(49,76)(48,76)(48,75)(47,75)(46,75)(46,74)(47,74)(47,73)(46,73)(46,72)(45,72)(45,73)
\psline(112,55)(112,56)(112,57)(113,57)(114,57)(114,56)(113,56)(113,55)(113,54)(114,54)(115,54)(115,53)(116,53)(116,54)(116,55)(115,55)(115,56)(116,56)(116,57)(117,57)(117,56)(117,55)(118,55)(118,54)
\psline(74,51)(75,51)(75,52)(74,52)(73,52)(72,52)(72,51)(71,51)(70,51)(69,51)(69,52)(68,52)(68,51)(67,51)(66,51)(66,52)(67,52)(67,53)(67,54)(68,54)(68,55)(69,55)(69,56)(70,56)
\psline(28,39)(27,39)(27,38)(27,37)(26,37)(26,38)(26,39)(26,40)(26,41)(26,42)(27,42)(27,43)(26,43)(26,44)(25,44)(24,44)(24,43)(23,43)(23,42)(22,42)(21,42)(21,43)(20,43)(20,44)
\psline(73,113)(73,112)(73,111)(73,110)(73,109)(73,108)(74,108)(74,109)(75,109)(75,110)(75,111)(75,112)(74,112)(74,113)(74,114)(74,115)(73,115)(73,114)(72,114)(72,115)(72,116)(71,116)(71,115)(71,114)
\psline(58,59)(57,59)(57,60)(58,60)(59,60)(59,59)(60,59)(61,59)(62,59)(62,58)(63,58)(63,59)(64,59)(64,60)(63,60)(62,60)(62,61)(61,61)(61,60)(60,60)(60,61)(59,61)(58,61)(58,62)
\psline(98,24)(99,24)(99,23)(98,23)(98,22)(98,21)(98,20)(98,19)(99,19)(99,18)(100,18)(100,19)(100,20)(101,20)(101,21)(102,21)(102,22)(101,22)(100,22)(100,23)(101,23)(102,23)(103,23)(104,23)
\psline(97,27)(98,27)(99,27)(99,26)(99,25)(98,25)(98,26)(97,26)(97,25)(96,25)(96,24)(95,24)(95,25)(95,26)(96,26)(96,27)(95,27)(95,28)(94,28)(94,29)(93,29)(93,28)(93,27)(93,26)
\psline(91,121)(91,122)(90,122)(90,123)(89,123)(88,123)(88,124)(88,125)(88,126)(87,126)(87,125)(86,125)(85,125)(85,126)(84,126)(84,127)(84,128)(83,128)(83,127)(82,127)(82,126)(83,126)(83,125)(83,124)
\psline(13,70)(12,70)(11,70)(11,71)(11,72)(10,72)(10,71)(10,70)(9,70)(8,70)(8,71)(9,71)(9,72)(9,73)(8,73)(8,72)(7,72)(7,73)(6,73)(6,74)(5,74)(5,75)(4,75)(4,74)
\psline(118,130)(119,130)(119,131)(118,131)(117,131)(117,132)(117,133)(117,134)(116,134)(116,133)(116,132)(116,131)(116,130)(116,129)(116,128)(116,127)(115,127)(115,128)(114,128)(114,129)(114,130)(113,130)(112,130)(111,130)
\psline(6,18)(5,18)(4,18)(4,19)(4,20)(4,21)(3,21)(2,21)(2,20)(3,20)(3,19)(3,18)(3,17)(4,17)(4,16)(4,15)(5,15)(5,14)(5,13)(4,13)(4,12)(5,12)(5,11)(4,11)
\psline(22,120)(21,120)(21,119)(22,119)(23,119)(23,120)(24,120)(24,119)(25,119)(26,119)(26,120)(26,121)(25,121)(25,122)(24,122)(24,121)(23,121)(23,122)(22,122)(22,121)(21,121)(20,121)(19,121)(19,122)
\psline(114,98)(113,98)(113,99)(113,100)(113,101)(112,101)(111,101)(111,100)(111,99)(110,99)(109,99)(108,99)(107,99)(107,100)(106,100)(105,100)(105,99)(104,99)(103,99)(102,99)(101,99)(101,100)(101,101)(102,101)
\psline(77,66)(78,66)(78,67)(78,68)(77,68)(76,68)(76,67)(76,66)(76,65)(75,65)(75,64)(74,64)(74,63)(73,63)(73,62)(74,62)(74,61)(74,60)(73,60)(73,59)(72,59)(72,60)(71,60)(70,60)
\psline(1,112)(1,113)(2,113)(3,113)(3,112)(3,111)(2,111)(1,111)(0,111)(0,110)(1,110)(2,110)(2,109)(1,109)(1,108)
\psline(136,108)(136,107)(136,106)
\psline(133,8)(132,8)(131,8)(131,9)(131,10)(131,11)(130,11)(129,11)(129,10)(130,10)(130,9)(129,9)(128,9)(127,9)(127,10)(126,10)(125,10)(125,11)(125,12)(124,12)(124,11)(123,11)(123,12)(123,13)
\psline(40,75)(41,75)(41,76)(41,77)(42,77)(42,76)(42,75)(43,75)(44,75)(44,74)(43,74)(42,74)(42,73)(43,73)(43,72)(44,72)(44,71)(43,71)(42,71)(41,71)(40,71)(40,72)(40,73)(41,73)
\psline(44,103)(44,102)(45,102)(46,102)(46,103)(47,103)(47,102)(47,101)(46,101)(45,101)(45,100)(46,100)(47,100)(47,99)(47,98)(46,98)(46,99)(45,99)(45,98)(45,97)(45,96)(45,95)(46,95)(46,96)
\psline(118,94)(119,94)(119,95)(118,95)(118,96)(119,96)(119,97)(120,97)(121,97)(121,96)(121,95)(121,94)(122,94)(122,93)(123,93)(124,93)(124,92)(123,92)(123,91)(124,91)(125,91)(125,90)(126,90)(126,89)
\psline(132,80)(132,79)(132,78)(131,78)(131,79)(130,79)(129,79)(129,78)(129,77)(129,76)(129,75)(130,75)(131,75)(131,76)(132,76)(132,77)(133,77)(133,76)(134,76)(135,76)(135,75)(135,74)(134,74)(133,74)
\psline(59,112)(59,113)(60,113)(60,112)(61,112)(62,112)(63,112)(63,111)(62,111)(61,111)(60,111)(59,111)(58,111)(58,112)(57,112)(57,113)(58,113)(58,114)(57,114)(57,115)(58,115)(59,115)(59,114)(60,114)
\psline(48,31)(47,31)(47,30)(47,29)(48,29)(48,30)(49,30)(49,31)(50,31)(50,32)(49,32)(48,32)(47,32)(46,32)(46,31)(46,30)(46,29)(45,29)(45,30)(45,31)(45,32)(44,32)(44,33)(43,33)
\psline(55,66)(55,67)(56,67)(57,67)(57,68)(58,68)(58,69)(57,69)(57,70)(57,71)(58,71)(58,70)(59,70)(60,70)(61,70)(61,71)(62,71)(62,72)(63,72)(63,71)(64,71)(65,71)(66,71)(66,72)
\psline(126,115)(126,116)(125,116)(125,117)(126,117)(127,117)(128,117)(129,117)(129,118)(130,118)(131,118)(132,118)(132,117)(131,117)(131,116)(130,116)(129,116)(128,116)(127,116)(127,115)(127,114)(126,114)(126,113)(125,113)
\psline(105,107)(105,108)(106,108)(106,107)(107,107)(107,108)(108,108)(108,109)(108,110)(107,110)(107,109)(106,109)(106,110)(105,110)(105,109)(104,109)(103,109)(103,110)(102,110)(101,110)(100,110)(100,111)(99,111)(99,110)
\psline(40,31)(41,31)(41,30)(41,29)(40,29)(39,29)(39,30)(39,31)(39,32)(39,33)(39,34)(39,35)(38,35)(37,35)(37,34)(37,33)(36,33)(36,32)(36,31)(35,31)(35,32)(35,33)(34,33)(33,33)
\psline(104,31)(104,30)(103,30)(103,31)(102,31)(102,30)(102,29)(101,29)(101,30)(101,31)(101,32)(102,32)(103,32)(104,32)(104,33)(103,33)(102,33)(102,34)(103,34)(104,34)(104,35)(105,35)(105,36)(106,36)
\psline(106,12)(106,11)(105,11)(105,12)(105,13)(105,14)(105,15)(105,16)(105,17)(104,17)(104,16)(103,16)(103,17)(103,18)(102,18)(102,19)(103,19)(103,20)(103,21)(104,21)(105,21)(105,20)(104,20)(104,19)
\psline(56,59)(55,59)(55,58)(54,58)(53,58)(52,58)(52,57)(51,57)(51,56)(51,55)(52,55)(52,54)(53,54)(53,55)(54,55)(54,56)(53,56)(53,57)(54,57)(55,57)(56,57)(56,56)(57,56)(57,57)
\psline(21,116)(22,116)(22,115)(23,115)(24,115)(24,114)(23,114)(22,114)(22,113)(21,113)(21,114)(21,115)(20,115)(19,115)(18,115)(18,116)(19,116)(20,116)(20,117)(19,117)(19,118)(20,118)(21,118)(22,118)
\psline(32,38)(31,38)(31,37)(32,37)(32,36)(32,35)(31,35)(30,35)(29,35)(28,35)(28,36)(28,37)(28,38)(29,38)(29,37)(30,37)(30,38)(30,39)(29,39)(29,40)(29,41)(30,41)(31,41)(32,41)
\psline(17,88)(17,89)(17,90)(16,90)(16,91)(16,92)(15,92)(14,92)(14,91)(15,91)(15,90)(15,89)(16,89)(16,88)(15,88)(15,87)(15,86)(15,85)(14,85)(14,84)(13,84)(13,83)(14,83)(15,83)
\psline(37,103)(38,103)(39,103)(40,103)(40,102)(39,102)(39,101)(38,101)(38,100)(37,100)(37,101)(37,102)(36,102)(36,101)(36,100)(36,99)(36,98)(37,98)(37,99)(38,99)(38,98)(38,97)(38,96)(38,95)
\psline(86,126)(86,127)(87,127)(88,127)(89,127)(90,127)(90,128)(89,128)(88,128)(87,128)(87,129)(86,129)(86,130)(86,131)(87,131)(87,130)(88,130)(88,129)(89,129)(89,130)(90,130)(91,130)(92,130)(92,129)
\psline(118,36)(118,35)(118,34)(117,34)(117,35)(117,36)(117,37)(116,37)(116,36)(116,35)(116,34)(115,34)(115,33)(116,33)(116,32)(115,32)(114,32)(114,33)(113,33)(113,32)(113,31)(113,30)(113,29)(113,28)
\psline(17,125)(18,125)(18,126)(18,127)(17,127)(16,127)(16,128)(16,129)(15,129)(15,128)(14,128)(13,128)(13,127)(14,127)(14,126)(15,126)(15,125)(15,124)(15,123)(15,122)(15,121)(15,120)(16,120)(16,121)
\psline(24,32)(24,33)(25,33)(25,34)(24,34)(23,34)(23,33)(23,32)(23,31)(22,31)(22,32)(21,32)(21,31)(20,31)(20,30)(21,30)(21,29)(20,29)(20,28)(21,28)(21,27)(21,26)(20,26)(19,26)
\psline(113,120)(113,121)(112,121)(112,122)(112,123)(111,123)(111,124)(112,124)(113,124)(113,123)(114,123)(114,124)(114,125)(113,125)(113,126)(113,127)(113,128)(112,128)(111,128)(110,128)(110,127)(109,127)(108,127)(108,128)
\psline(41,8)(41,9)(42,9)(42,10)(41,10)(40,10)(40,9)(40,8)(40,7)(40,6)(40,5)(40,4)(39,4)(39,3)(40,3)(41,3)(41,4)(42,4)(42,5)(43,5)(43,4)(44,4)(45,4)(46,4)
\psline(77,15)(77,14)(78,14)(78,15)(78,16)(77,16)(77,17)(78,17)(78,18)(78,19)(78,20)(79,20)(79,19)(79,18)(80,18)(81,18)(81,19)(80,19)(80,20)(81,20)(81,21)(80,21)(80,22)(81,22)
\psline(85,113)(85,114)(86,114)(87,114)(88,114)(88,113)(87,113)(87,112)(87,111)(86,111)(86,110)(85,110)(85,111)(84,111)(83,111)(83,110)(83,109)(84,109)(85,109)(86,109)(86,108)(86,107)(85,107)(85,108)
\psline(91,10)(90,10)(90,11)(91,11)(91,12)(92,12)(92,11)(92,10)(93,10)(93,9)(92,9)(92,8)(93,8)(93,7)(94,7)(95,7)(96,7)(96,8)(96,9)(96,10)(97,10)(97,11)(96,11)(96,12)
\psline(92,65)(92,66)(93,66)(93,65)(94,65)(95,65)(95,64)(96,64)(96,63)(97,63)(98,63)(98,64)(97,64)(97,65)(98,65)(99,65)(99,66)(98,66)(97,66)(96,66)(96,67)(96,68)(97,68)(98,68)
\psline(11,67)(12,67)(12,66)(12,65)(11,65)(11,64)(10,64)(9,64)(8,64)(8,65)(9,65)(9,66)(8,66)(7,66)(7,67)(7,68)(8,68)(8,69)(9,69)(10,69)(10,68)(11,68)(12,68)(12,69)
\psline(19,20)(19,21)(19,22)(18,22)(17,22)(17,23)(18,23)(18,24)(17,24)(17,25)(18,25)(18,26)(18,27)(18,28)(19,28)(19,29)(19,30)(19,31)(18,31)(17,31)(16,31)(15,31)(15,32)(15,33)
\psline(127,127)(126,127)(126,128)(126,129)(127,129)(127,128)(128,128)(128,127)(128,126)(127,126)(127,125)(128,125)(129,125)(129,124)(130,124)(131,124)(131,125)(130,125)(130,126)(129,126)(129,127)(129,128)(129,129)(129,130)
\psline(110,125)(109,125)(109,124)(109,123)(110,123)(110,122)(109,122)(108,122)(108,123)(107,123)(106,123)(105,123)(105,122)(104,122)(104,121)(104,120)(104,119)(103,119)(103,120)(102,120)(102,119)(101,119)(101,120)(100,120)
\psline(125,51)(125,50)(124,50)(124,49)(123,49)(123,48)(123,47)(124,47)(124,48)(125,48)(125,49)(126,49)(126,48)(126,47)(127,47)(128,47)(129,47)(130,47)(130,48)(131,48)(131,49)(132,49)(132,48)(133,48)
\psline(24,36)(23,36)(22,36)(22,37)(21,37)(20,37)(20,38)(21,38)(21,39)(22,39)(23,39)(23,40)(23,41)(22,41)(22,40)(21,40)(20,40)(20,41)(20,42)(19,42)(19,43)(19,44)(19,45)(19,46)
\psline(30,68)(31,68)(31,67)(30,67)(30,66)(31,66)(31,65)(32,65)(32,64)(33,64)(34,64)(34,63)(35,63)(35,64)(35,65)(34,65)(33,65)(33,66)(32,66)(32,67)(33,67)(34,67)(34,66)(35,66)
\psline(40,26)(41,26)(41,25)(41,24)(41,23)(42,23)(43,23)(43,22)(44,22)(45,22)(45,21)(45,20)(44,20)(44,21)(43,21)(43,20)(42,20)(42,21)(41,21)(41,20)(41,19)(41,18)(40,18)(40,17)
\psline(7,34)(6,34)(5,34)(5,35)(6,35)(6,36)(7,36)(7,37)(7,38)(6,38)(6,37)(5,37)(4,37)(3,37)(3,38)(2,38)(1,38)
\psline(136,38)(136,37)
\psline(0,36)(0,35)(1,35)
\psline(50,1)(50,2)(51,2)(51,1)(52,1)(53,1)(54,1)
\psline(55,0)(56,0)(57,0)
\psline(58,136)(59,136)
\psline(62,0)(62,1)(63,1)
\psline(22,63)(22,64)(23,64)(24,64)(25,64)(25,63)(24,63)(24,62)(23,62)(22,62)(21,62)(21,63)(20,63)(19,63)(19,62)(19,61)(20,61)(20,60)(21,60)(21,59)(21,58)(22,58)(22,57)(21,57)
\psline(34,94)(33,94)(32,94)(31,94)(30,94)(30,93)(30,92)(29,92)(28,92)(28,93)(27,93)(26,93)(25,93)(24,93)(23,93)(23,92)(24,92)(24,91)(25,91)(26,91)(26,92)(27,92)(27,91)(28,91)
\psline(132,98)(132,97)(131,97)(131,96)(132,96)(132,95)(133,95)(133,96)(134,96)(134,95)(134,94)(135,94)
\psline(136,93)(136,92)(136,91)
\psline(0,90)(0,91)(0,92)(1,92)(1,93)
\psline(135,106)(134,106)(133,106)(132,106)(131,106)(130,106)(130,105)(129,105)(128,105)(128,106)(127,106)(126,106)(125,106)(125,105)(126,105)(126,104)(127,104)(127,103)(126,103)(126,102)(126,101)(125,101)(125,100)(125,99)
\psline(129,65)(129,66)(128,66)(128,67)(127,67)(127,66)(127,65)(128,65)(128,64)(127,64)(127,63)(126,63)(125,63)(125,64)(126,64)(126,65)(126,66)(126,67)(126,68)(126,69)(125,69)(125,70)(124,70)(124,71)
\psline(22,66)(22,67)(22,68)(23,68)(23,69)(23,70)(24,70)(24,71)(25,71)(26,71)(27,71)(27,72)(28,72)(29,72)(29,73)(29,74)(28,74)(28,73)(27,73)(26,73)(26,72)(25,72)(24,72)(24,73)
\psline(127,60)(128,60)(129,60)(129,59)(128,59)(128,58)(128,57)(129,57)(130,57)(130,56)(130,55)(130,54)(129,54)(129,53)(128,53)(127,53)(127,52)(126,52)(126,51)(126,50)(127,50)(128,50)(128,49)(127,49)
\psline(109,21)(108,21)(108,22)(109,22)(109,23)(108,23)(108,24)(109,24)(110,24)(110,23)(110,22)(111,22)(111,21)(112,21)(112,22)(113,22)(113,23)(112,23)(111,23)(111,24)(111,25)(110,25)(109,25)(108,25)
\psline(22,80)(21,80)(20,80)(20,79)(21,79)(22,79)(23,79)(24,79)(25,79)(25,78)(26,78)(27,78)(27,77)(27,76)(28,76)(28,77)(28,78)(29,78)(29,79)(28,79)(27,79)(26,79)(26,80)(26,81)
\psline(18,60)(19,60)(19,59)(20,59)(20,58)(20,57)(19,57)(19,58)(18,58)(18,59)(17,59)(17,60)(17,61)(17,62)(17,63)(17,64)(17,65)(16,65)(16,66)(16,67)(15,67)(15,66)(15,65)(14,65)
\psline(33,133)(34,133)(35,133)(36,133)(36,132)(35,132)(35,131)(34,131)(34,130)(34,129)(33,129)(33,130)(33,131)(32,131)(32,130)(32,129)(31,129)(31,128)(32,128)(33,128)(33,127)(33,126)(34,126)(35,126)
\psline(95,121)(95,120)(95,119)(94,119)(94,118)(93,118)(93,119)(92,119)(92,120)(93,120)(93,121)(92,121)(92,122)(92,123)(91,123)(91,124)(90,124)(89,124)(89,125)(89,126)(90,126)(90,125)(91,125)(92,125)
\psline(81,110)(80,110)(79,110)(79,109)(80,109)(80,108)(81,108)(81,109)(82,109)(82,110)(82,111)(81,111)(81,112)(82,112)(82,113)(83,113)(83,114)(82,114)(82,115)(82,116)(81,116)(81,115)(81,114)(80,114)
\psline(77,48)(77,47)(77,46)(76,46)(75,46)(75,47)(75,48)(74,48)(73,48)(73,47)(73,46)(73,45)(73,44)(74,44)(74,45)(75,45)(76,45)(77,45)(78,45)(78,44)(79,44)(79,45)(79,46)(79,47)
\psline(128,51)(128,52)(129,52)(129,51)(130,51)(130,52)(131,52)(132,52)(132,53)(133,53)(133,52)(134,52)(134,51)(133,51)(132,51)(132,50)(133,50)(134,50)(135,50)(135,49)(134,49)(134,48)(135,48)
\psline(93,117)(92,117)(91,117)(90,117)(90,116)(89,116)(89,117)(89,118)(88,118)(87,118)(86,118)(85,118)(85,117)(86,117)(86,116)(86,115)(85,115)(84,115)(84,116)(84,117)(84,118)(83,118)(83,117)(82,117)
\psline(75,121)(74,121)(73,121)(73,120)(73,119)(72,119)(72,118)(71,118)(71,119)(71,120)(71,121)(72,121)(72,122)(71,122)(70,122)(69,122)(69,123)(69,124)(68,124)(67,124)(67,125)(66,125)(65,125)(65,124)
\psline(68,37)(68,38)(67,38)(67,39)(66,39)(66,40)(66,41)(66,42)(65,42)(64,42)(64,43)(65,43)(66,43)(66,44)(67,44)(68,44)(68,45)(67,45)(66,45)(65,45)(65,44)(64,44)(63,44)(63,43)
\psline(71,53)(71,52)(70,52)(70,53)(70,54)(71,54)(71,55)(72,55)(73,55)(73,56)(72,56)(72,57)(71,57)(71,58)(72,58)(73,58)(73,57)(74,57)(74,56)(74,55)(74,54)(75,54)(76,54)(76,53)
\psline(108,124)(107,124)(107,125)(106,125)(106,126)(106,127)(107,127)(107,128)(107,129)(107,130)(106,130)(105,130)(105,131)(104,131)(104,132)(104,133)(105,133)(105,134)(105,135)(104,135)(103,135)(103,136)
\psline(115,8)(115,9)(115,10)(116,10)(116,9)(116,8)(117,8)(118,8)(119,8)(119,9)(119,10)(119,11)(119,12)(119,13)(120,13)(120,12)(120,11)(121,11)(121,10)(122,10)(122,11)(122,12)(122,13)(122,14)
\psline(98,103)(98,102)(97,102)(97,103)(97,104)(98,104)(98,105)(98,106)(99,106)(100,106)(101,106)(102,106)(102,107)(102,108)(103,108)(104,108)(104,107)(103,107)(103,106)(103,105)(104,105)(104,106)(105,106)(105,105)
\psline(84,132)(84,133)(83,133)(82,133)(81,133)(81,132)(82,132)(83,132)(83,131)(82,131)(82,130)(83,130)(83,129)(82,129)(82,128)(81,128)(81,129)(81,130)(80,130)(80,129)(80,128)(79,128)(79,127)(80,127)
\psline(0,11)(1,11)(1,10)(1,9)(2,9)(2,10)(2,11)(2,12)(2,13)(2,14)(3,14)(3,15)(3,16)(2,16)(2,15)(1,15)(1,14)(1,13)(1,12)(0,12)(0,13)
\psline(40,69)(40,68)(41,68)(41,69)(42,69)(42,70)(41,70)(40,70)(39,70)(39,69)(38,69)(38,70)(38,71)(38,72)(37,72)(36,72)(36,71)(37,71)(37,70)(37,69)(36,69)(35,69)(34,69)(33,69)
\psline(35,80)(35,79)(35,78)(34,78)(34,77)(33,77)(33,78)(32,78)(31,78)(30,78)(30,79)(30,80)(29,80)(28,80)(28,81)(29,81)(29,82)(29,83)(28,83)(28,82)(27,82)(26,82)(26,83)(25,83)
\psline(122,114)(122,115)(123,115)(123,114)(123,113)(123,112)(124,112)(125,112)(126,112)(126,111)(126,110)(125,110)(125,111)(124,111)(123,111)(123,110)(124,110)(124,109)(124,108)(123,108)(122,108)(122,109)(121,109)(120,109)
\psline(0,97)(0,98)(1,98)(2,98)(2,99)(3,99)(3,100)(3,101)(4,101)(4,102)(3,102)(3,103)(2,103)(1,103)(1,104)(1,105)(1,106)(0,106)(0,107)(1,107)(2,107)(2,106)
\psline(114,6)(114,5)(113,5)(113,6)(112,6)(112,7)(111,7)(110,7)(110,6)(110,5)(110,4)(111,4)(112,4)(112,3)(113,3)(113,4)(114,4)(114,3)(114,2)(113,2)(113,1)(113,0)(114,0)
\psline(75,10)(76,10)(76,9)(75,9)(75,8)(74,8)(74,9)(74,10)(74,11)(73,11)(73,10)(72,10)(71,10)(71,11)(70,11)(70,12)(69,12)(68,12)(68,13)(69,13)(69,14)(70,14)(70,15)(69,15)
\psline(67,96)(67,97)(66,97)(66,96)(66,95)(65,95)(65,94)(64,94)(64,93)(65,93)(65,92)(65,91)(66,91)(67,91)(68,91)(68,92)(67,92)(67,93)(67,94)(68,94)(68,93)(69,93)(69,92)(70,92)
\psline(55,111)(54,111)(54,110)(55,110)(56,110)(56,111)(56,112)(55,112)(54,112)(53,112)(52,112)(51,112)(51,113)(51,114)(52,114)(52,115)(53,115)(53,114)(53,113)(54,113)(55,113)(56,113)(56,114)(56,115)
\psline(59,36)(60,36)(60,37)(59,37)(59,38)(60,38)(61,38)(61,39)(61,40)(62,40)(62,39)(63,39)(63,40)(63,41)(63,42)(62,42)(62,41)(61,41)(60,41)(60,40)(59,40)(59,41)(58,41)(58,42)
\psline(55,62)(55,63)(56,63)(57,63)(58,63)(59,63)(60,63)(61,63)(61,62)(62,62)(63,62)(63,63)(62,63)(62,64)(63,64)(64,64)(64,63)(65,63)(65,64)(65,65)(65,66)(65,67)(66,67)(67,67)
\psline(116,67)(116,66)(116,65)(117,65)(117,64)(118,64)(118,63)(118,62)(119,62)(119,63)(119,64)(119,65)(118,65)(118,66)(118,67)(118,68)(118,69)(118,70)(119,70)(120,70)(121,70)(122,70)(122,69)(121,69)
\psline(121,41)(121,42)(121,43)(120,43)(120,42)(120,41)(120,40)(119,40)(118,40)(118,39)(117,39)(117,38)(118,38)(118,37)(119,37)(119,36)(120,36)(120,35)(120,34)(120,33)(119,33)(119,32)(118,32)(117,32)
\psline(10,73)(11,73)(11,74)(11,75)(10,75)(10,76)(11,76)(11,77)(12,77)(12,76)(13,76)(13,77)(14,77)(14,78)(15,78)(16,78)(16,79)(16,80)(16,81)(15,81)(14,81)(13,81)(12,81)(11,81)
\psline(5,77)(6,77)(6,78)(5,78)(5,79)(4,79)(3,79)(2,79)(2,78)(2,77)(3,77)(3,78)(4,78)(4,77)(4,76)(5,76)(6,76)(6,75)(7,75)(7,74)(8,74)(8,75)(9,75)(9,76)
\psline(91,40)(92,40)(93,40)(94,40)(95,40)(96,40)(96,39)(96,38)(95,38)(95,37)(94,37)(93,37)(92,37)(91,37)(91,38)(90,38)(90,39)(89,39)(89,40)(89,41)(89,42)(89,43)(88,43)(88,44)
\psline(47,86)(48,86)(48,85)(49,85)(49,86)(50,86)(50,87)(49,87)(49,88)(48,88)(48,89)(49,89)(50,89)(50,88)(51,88)(51,87)(52,87)(52,86)(52,85)(52,84)(52,83)(51,83)(50,83)(50,82)
\psline(11,99)(10,99)(10,98)(11,98)(11,97)(12,97)(12,96)(11,96)(11,95)(10,95)(10,94)(11,94)(11,93)(11,92)(11,91)(10,91)(10,92)(9,92)(9,93)(8,93)(8,92)(7,92)(7,93)(6,93)
\psline(70,70)(70,69)(71,69)(72,69)(72,70)(73,70)(73,71)(73,72)(72,72)(72,73)(71,73)(71,72)(71,71)(70,71)(70,72)(70,73)(69,73)(69,74)(68,74)(68,75)(67,75)(67,76)(66,76)(66,75)
\psline(65,40)(65,39)(64,39)(64,38)(65,38)(65,37)(65,36)(65,35)(65,34)(64,34)(63,34)(63,35)(64,35)(64,36)(63,36)(62,36)(62,35)(62,34)(61,34)(60,34)(60,35)(59,35)(58,35)(58,34)
\psline(49,79)(49,80)(49,81)(50,81)(50,80)(50,79)(51,79)(51,80)(52,80)(53,80)(54,80)(54,81)(55,81)(56,81)(56,80)(57,80)(57,79)(57,78)(58,78)(58,77)(58,76)(57,76)(57,75)(58,75)
\psline(8,134)(8,133)(8,132)(7,132)(6,132)(6,131)(5,131)(5,130)(5,129)(6,129)(6,130)(7,130)(8,130)(9,130)(9,131)(10,131)(10,130)(11,130)(12,130)(12,129)(11,129)(10,129)(9,129)(9,128)
\psline(61,65)(62,65)(62,66)(61,66)(61,67)(62,67)(62,68)(63,68)(63,67)(63,66)(63,65)(64,65)(64,66)(64,67)(64,68)(65,68)(65,69)(64,69)(64,70)(65,70)(66,70)(67,70)(67,71)(68,71)
\psline(80,90)(80,89)(80,88)(79,88)(78,88)(78,87)(78,86)(79,86)(79,87)(80,87)(80,86)(80,85)(81,85)(82,85)(83,85)(84,85)(84,84)(84,83)(85,83)(86,83)(87,83)(87,84)(88,84)(89,84)
\psline(26,36)(27,36)(27,35)(26,35)(26,34)(26,33)(26,32)(27,32)(27,31)(27,30)(26,30)(26,29)(27,29)(28,29)(29,29)(29,28)(29,27)(28,27)(28,28)(27,28)(27,27)(26,27)(26,26)(27,26)
\psline(133,45)(132,45)(132,44)(133,44)(134,44)(135,44)(135,45)(134,45)(134,46)(134,47)(133,47)(132,47)(132,46)(131,46)(131,45)(130,45)(129,45)(129,44)(130,44)(130,43)(129,43)(128,43)(128,44)(128,45)
\psline(77,0)(78,0)(79,0)(80,0)
\psline(81,136)(81,135)(80,135)(79,135)(78,135)(78,134)(79,134)(80,134)(81,134)(82,134)(82,135)
\psline(82,0)(83,0)(84,0)(84,1)(85,1)
\psline(45,7)(45,6)(44,6)(43,6)(43,7)(42,7)(42,8)(43,8)(43,9)(44,9)(44,8)(45,8)(46,8)(46,9)(45,9)(45,10)(46,10)(47,10)(47,11)(46,11)(45,11)(45,12)(44,12)(44,11)
\psline(23,60)(22,60)(22,61)(23,61)(24,61)(25,61)(26,61)(26,60)(27,60)(27,59)(26,59)(25,59)(25,58)(24,58)(23,58)(23,57)(24,57)(24,56)(25,56)(26,56)(26,55)(25,55)(25,54)(26,54)
\psline(116,104)(115,104)(115,105)(116,105)(116,106)(115,106)(115,107)(115,108)(114,108)(113,108)(112,108)(112,109)(111,109)(111,110)(112,110)(112,111)(111,111)(111,112)(111,113)(110,113)(110,114)(111,114)(111,115)(112,115)
\psline(46,125)(45,125)(44,125)(44,126)(45,126)(46,126)(46,127)(47,127)(47,126)(48,126)(48,127)(48,128)(48,129)(47,129)(47,130)(48,130)(48,131)(48,132)(47,132)(47,131)(46,131)(45,131)(45,130)(44,130)
\psline(110,21)(110,20)(109,20)(109,19)(109,18)(109,17)(108,17)(107,17)(107,16)(108,16)(108,15)(109,15)(109,14)(108,14)(108,13)(108,12)(108,11)(107,11)(107,10)(108,10)(108,9)(108,8)(108,7)(109,7)
\psline(28,106)(28,107)(27,107)(27,108)(28,108)(29,108)(29,109)(30,109)(31,109)(32,109)(32,108)(33,108)(34,108)(34,109)(33,109)(33,110)(34,110)(35,110)(36,110)(37,110)(38,110)(38,111)(37,111)(37,112)
\psline(69,80)(70,80)(71,80)(72,80)(72,81)(72,82)(73,82)(73,83)(73,84)(73,85)(72,85)(72,86)(73,86)(73,87)(72,87)(72,88)(71,88)(71,89)(71,90)(72,90)(72,89)(73,89)(74,89)(74,88)
\psline(108,44)(109,44)(109,43)(110,43)(110,44)(110,45)(110,46)(110,47)(111,47)(111,46)(111,45)(112,45)(112,44)(113,44)(114,44)(114,43)(115,43)(115,44)(116,44)(117,44)(118,44)(119,44)(120,44)(120,45)
\psline(100,78)(99,78)(98,78)(98,77)(97,77)(97,76)(96,76)(96,77)(96,78)(95,78)(95,79)(94,79)(93,79)(92,79)(91,79)(91,78)(92,78)(92,77)(93,77)(93,76)(93,75)(94,75)(94,76)(95,76)
\psline(30,75)(30,74)(31,74)(31,73)(32,73)(32,74)(32,75)(33,75)(33,76)(34,76)(34,75)(34,74)(33,74)(33,73)(34,73)(34,72)(34,71)(35,71)(35,70)(34,70)(33,70)(32,70)(31,70)(30,70)
\psline(88,78)(88,79)(89,79)(90,79)(90,80)(89,80)(88,80)(87,80)(87,81)(87,82)(88,82)(88,81)(89,81)(89,82)(90,82)(90,81)(91,81)(92,81)(92,82)(91,82)(91,83)(90,83)(90,84)(90,85)
\psline(127,6)(126,6)(126,5)(125,5)(125,4)(124,4)(123,4)(122,4)(122,3)(122,2)(121,2)(120,2)(120,1)(120,0)
\psline(119,136)(118,136)(118,135)(118,134)(119,134)(119,135)(120,135)(120,134)(121,134)
\psline(20,135)(20,134)(19,134)(18,134)(18,133)(18,132)(18,131)(18,130)(17,130)(17,129)(18,129)(18,128)(19,128)(19,127)(20,127)(20,126)(20,125)(20,124)(20,123)(21,123)(22,123)(23,123)(24,123)(24,124)
\psline(107,120)(108,120)(108,121)(107,121)(107,122)(106,122)(106,121)(106,120)(106,119)(105,119)(105,118)(104,118)(104,117)(103,117)(103,116)(102,116)(101,116)(101,117)(101,118)(100,118)(99,118)(98,118)(98,119)(99,119)
\psline(53,132)(53,131)(53,130)(52,130)(52,131)(52,132)(51,132)(51,133)(52,133)(53,133)(54,133)(55,133)(55,134)(54,134)(54,135)(53,135)(53,134)(52,134)(52,135)(52,136)(51,136)(51,135)(51,134)(50,134)
\psline(115,85)(116,85)(117,85)(118,85)(118,84)(119,84)(119,83)(120,83)(120,84)(120,85)(119,85)(119,86)(120,86)(120,87)(120,88)(121,88)(121,87)(121,86)(121,85)(121,84)(122,84)(123,84)(124,84)(125,84)
\psline(30,88)(30,87)(31,87)(31,88)(31,89)(30,89)(30,90)(30,91)(31,91)(32,91)(32,90)(32,89)(32,88)(33,88)(33,89)(34,89)(35,89)(35,88)(36,88)(36,89)(37,89)(37,90)(36,90)(35,90)
\psline(63,131)(63,130)(63,129)(62,129)(61,129)(61,130)(60,130)(60,131)(61,131)(61,132)(62,132)(63,132)(63,133)(64,133)(65,133)(66,133)(66,134)(66,135)(67,135)(67,136)
\psline(68,0)(68,1)(67,1)
\psline(43,31)(44,31)(44,30)(44,29)(43,29)(42,29)(42,30)(42,31)(42,32)(41,32)(40,32)(40,33)(40,34)(40,35)(40,36)(39,36)(39,37)(40,37)(41,37)(42,37)(42,38)(41,38)(41,39)(40,39)
\psline(123,38)(123,37)(122,37)(122,36)(121,36)(121,35)(121,34)(121,33)(121,32)(120,32)(120,31)(121,31)(122,31)(123,31)(124,31)(124,32)(124,33)(124,34)(124,35)(123,35)(123,34)(122,34)(122,33)(123,33)
\psline(113,46)(113,47)(114,47)(114,46)(114,45)(115,45)(115,46)(115,47)(115,48)(115,49)(115,50)(115,51)(115,52)(114,52)(113,52)(112,52)(112,51)(112,50)(111,50)(111,49)(110,49)(110,48)(109,48)(109,47)
\psline(77,77)(77,76)(77,75)(76,75)(76,76)(76,77)(76,78)(75,78)(75,79)(74,79)(74,78)(73,78)(73,79)(73,80)(73,81)(74,81)(74,82)(75,82)(75,83)(74,83)(74,84)(74,85)(75,85)(75,84)
\psline(79,27)(79,26)(80,26)(80,27)(80,28)(81,28)(81,27)(81,26)(81,25)(80,25)(79,25)(79,24)(80,24)(81,24)(82,24)(82,23)(83,23)(84,23)(84,24)(85,24)(85,23)(85,22)(86,22)(86,21)
\psline(52,81)(51,81)(51,82)(52,82)(53,82)(54,82)(55,82)(56,82)(57,82)(57,81)(58,81)(59,81)(60,81)(61,81)(62,81)(62,82)(63,82)(64,82)(65,82)(65,83)(64,83)(64,84)(64,85)(65,85)
  \end{pspicture}
\end{center}
	
	\vspace{-0.5cm}
	\caption{Example of a state of a two-dimensional multi-polymer system on the torus $\big(\Z / 135\Z\big)^2$ with $100$ polymers of length $25$ each.}\label{fig:example_configuration}
\end{figure}
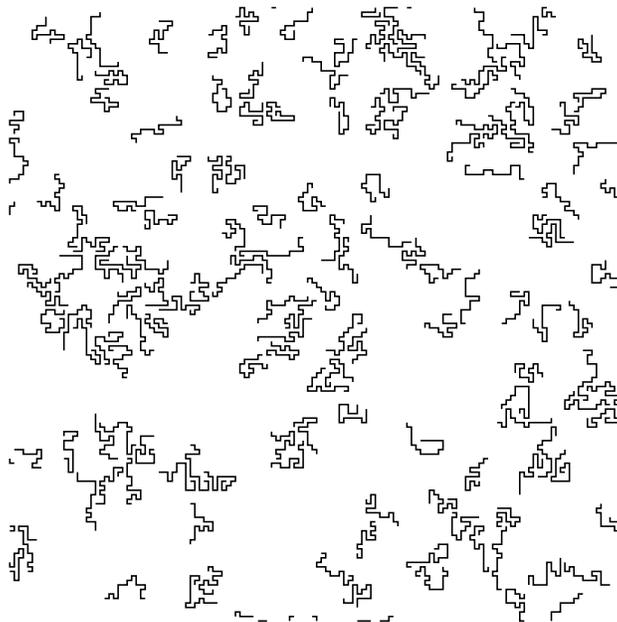

The state space of the \R algorithm is the set of all possible configurations of non-intersecting polymers in some dimension $d$. More precisely, there is an underlying $d$-dimensional grid, endowed with a cyclic structure on the edges, on which all the objects are considered. A polymer is then just an undirected self-avoiding path of given length, a path being a sequence of neighboring vertices on the underlying grid. The fact that polymers cannot intersect each other, and that they cannot intersect themselves, comes from a very natural physical constraint, namely that a point in space cannot be occupied by two different particles. Figure~\ref{fig:example_configuration} shows an example of a two-dimensional multi-polymer system on a torus with $100$ polymers of length $25$ each.

One step of this algorithm consists in removing one polymer at random, and trying to re-grow it elsewhere. The growth may not be successful, in which case we put back the original polymer in the system, and we remain in the same state. But if we successfully grew a candidate polymer, by analogy with the Metropolis algorithm, we flip a coin to decide whether to add this polymer or to put back the original one. The goal of this coin-tossing is to compensate for the bias introduced during the growth of the candidate polymer in order to yield the right stationary distribution. Though it does not clearly appear at this point, we want to stress the fact that computing this probability of acceptance will require a fair amount of work. This is in sharp contrast with the Metropolis algorithm, for which it is straightforward to derive this quantity.

One of the main difficulties of the \R algorithm therefore lies in the growth of the new candidate polymer. Since polymers are self-avoiding paths, this problem is closely related to the generation of self-avoiding walks, for which an important amount of research has already been carried out~\cite{saw-book, saw}. Two differences make the situation slightly more complex in our case. First, we have several polymers here, with the constraint that they cannot intersect each other. Second, each step of the \R algorithm involves the attempt to grow a new polymer. In order for the algorithm to perform well, this growth needs to be done fast, even if it means that it fails often.
\\

The efficiency of the \R algorithm, and hence its interest compared to other algorithms, can be assessed by its ability to address a number of issues that clearly emerge from the above global description.

First, the \R algorithm must generate a state $S$ with probability arbitrarily close to $q(S)$, where $q$ is the distribution we want to sample from. We have seen that this will be done by constructing a Markov chain having $q$ as the stationary distribution. As in the Metropolis algorithm, once we have a candidate polymer, we are free to choose the probability of accepting it; we prove in this paper that the chain will be reversible for a suitable choice of the probability of acceptance. By doing so, we find that the probability of acceptance suggested in~\cite{Consta} is the right one. However, the same probability of acceptance fails to yield the right stationary distribution in some extended version of the algorithm, and this subtle mistake was not detected in the original paper.

Second, the algorithm must mix as quickly as possible, which means that stationarity has to be reached as fast as possible. Even on simple examples, determining the convergence rate of a Metropolis type algorithm is a very hard question, as is shown in~\cite{Metropolis}. Nevertheless, a rapidly mixing Markov chain requires both high chain construction and high acceptance rates. The \emph{chain construction rate} refers to the probability that the growth of a new candidate polymer is successful, while the \emph{acceptance rate} refers to the probability that, given a candidate polymer, it is accepted. We see from the above description of the \R algorithm that if one of these two rates is low, the algorithm will stay for a long time in the same state before changing, so that it is  very unlikely that it mixes rapidly.

These two quantities intrinsically depend on the generation of the new candidate polymer. For instance, we have seen that the rejection procedure of the Metropolis algorithm depends on the ratio $A(y,x)/A(x,y)$, with $A(x,y) = \pi(x)P(x,y)$: since the candidate $y$ for the new state is chosen with probability $P(x,y)$, the question is to know whether $P(x,\cdot)$ gives advantage to states that are more likely for $\pi$. If so, then this latter ratio will often be above one, yielding a high acceptance rate. This is especially clear if $P$ is chosen to be symmetric, in which case the acceptance probability is $\min(1, \pi(y)/\pi(x))$. In other words, a good algorithm favors candidate states that yield a good acceptance probability. As for the construction rate, a trade-off has to be made between the construction rate itself and the computational cost. Heuristically, the algorithm that generates a new polymer works by growing or recoiling one step at a time, until the polymer under construction has reached the desired length. In particular, this algorithm lives in the space of incomplete self-avoiding paths. This space is huge, especially if the dimension is large or if the polymer is long. The trade-off is therefore the following: if one allows the algorithm to visit all the space of incomplete self-avoiding paths, the construction rate will be the best possible, but the computational cost may be incredibly expensive. By truncating the set of incomplete self-avoiding paths which the algorithm is allowed to visit, one reduces this computational cost at the expense of reducing the construction rate as well.
\\

The \R algorithm has two salient features that yield both high construction and acceptance rates: the concepts of \emph{feeler} and of \emph{underlying graph}. Though these two concepts are combined in the \R algorithm, it is worth emphasizing that they can be considered and defined independently from one another. The idea of feeler makes it possible to define an algorithm that generates a polymer step by step by looking a few steps ahead. This way, the algorithm is able to detect dense areas where it becomes hard to grow the polymer, and, if necessary, to recoil from such regions. The feeler is characterized by a parameter $\ell$, the length of the feeler, that tells how far away the algorithm is allowed to look. Large values of $\ell$ yield a better construction rate, but a higher computational cost.

As we will see, the algorithm that attempts to grow a new polymer works by increasing or decreasing a partial polymer by only one vertex at a time. In particular, this algorithm can be defined on any oriented graph: all that matters is to know the potential growth directions, or equivalently, the neighbors of the endpoint of the current partial polymer. However, the properties of the algorithm, namely the computational cost and the construction and acceptance rates, are crucially affected by the graph on which it is run.

The set of underlying graphs is the class of graphs on which the growth procedure, that makes use of the concept of feeler, will be run. Intuitively, we see that the more neighbors are allowed at each step, the higher the computational cost. Initially, everything happens on a $d$-dimensional grid, in which every vertex has $Q=2d$ neighbors. An idea to lower the computational complexity is to impose vertices in an underlying graph to have the same out-degree $k < Q$, where $k$ is a parameter of our algorithm to be adjusted.

With these remarks in mind, a simple sketch of the \R algorithm is as follows. First, choose an old polymer and remove it from the system. Then generate at random an underlying graph, and try to grow a candidate polymer on this underlying graph using the idea of feeler. If the growth is not successful, then go back to the initial state. Otherwise, toss a coin to determine whether to accept or not this candidate. Clearly, the generation of an underlying graph conditions the further growth of the candidate polymer. In particular, the probability of acceptance depends on the underlying graph generated, which makes the connection with the auxiliary variable method. As we will see, delving into the details of every of these steps will require a lot of work. It is however very important to keep this simple sketch in mind.
\\

The rest of the paper is organized as follows. In Section~\ref{sec:framework}, we define the notations and the global framework, and give an overview of the concepts of underlying graph and feeler by comparing the \R algorithm to two simpler algorithms. Section~\ref{sec:rga} then explains the \R algorithm in details and computes the generating probabilities of the objects under consideration. We then prove in Section~\ref{sec:properties} that the Markov Chain run by the algorithm has the right stationary distribution, and find the correct probability of acceptance. Some considerations about the irreducibility of the algorithm are given in Section~\ref{sub:irreducibility}, and some possible extensions of the \R algorithm are studied in Section~\ref{sub:extensions}. Finally, Section~\ref{sec:implementation} is devoted to issues related to practical implementations of the algorithm.

\section{Framework}\label{sec:framework}

\subsection{General settings, notations}\label{sub:notations}

In all that follows, we are working on a finite $d$-dimensional lattice $\G$ in which each vertex has $Q = 2d$ neighbors. $\G$ will be referred to as the \textit{underlying lattice}. To make sure that the lattice is finite and that each vertex has exactly $Q$ neighbors, $\G$ is endowed with a torus structure. So if one thinks of $\G$ as a cube in dimension $d$, some vertices ``go around''. For $d = 1$, $\G$ is a circle, and for $d = 2$, it is a torus.

More formally, the set of vertices of $\G$ is $\left(\Z / a \Z\right)^d$ for some fixed $a$, and two vertices are neighbors in $\G$ if all their coordinates are equal, except for one in which they differ by $1$ or $-1$ (in $\Z / a\Z$).
\\

On this lattice, our system consists of $N$ polymers of fixed length. A polymer $(v_1, \ldots, v_L)$ is a self-avoiding path of length $L \geq 1$, the length being the number of vertices. Moreover, the same physical constraints that impose the polymers to be self-avoiding impose that two different polymers cannot intersect. We denote by $\mathcal S$ the state space of all the possible configurations, so $\mathcal S$ can be written:
\begin{align*}
  \mathcal S = \{(C_k), k=1\ldots N, C_k \textrm{ is a self-avoiding} &\textrm{ path of length $L$}\\ & \textrm{and } C_i \cap C_j = \emptyset \textrm{ for } i \neq j \}
\end{align*}

Figure~\ref{fig:example_configuration} shows a particular state of a system with $N=100$ polymers of length $L=25$ each and in dimension $d=2$, so that the underlying lattice is actually a torus. Note that some polymers, which are in one piece on the torus, are disconnected in such a plane representation.

When $L=1$, polymers are reduced to a vertex; for $L=2$, the polymers are actually called dimers. A fair amount of literature exists on the so-called dimer-monomer problem, but none of them covers our case. There is a total of $a^d$ vertices, and the polymers occupy $NL$ vertices: it is easy to see that as soon as $NL \leq a^d$, it is possible to cover the underlying lattice with the specified polymers. Indeed, there clearly exists a self-avoiding path of length $a^d$ that covers $\G$: this is just a matter of suitably enumerating $\big(\Z / a\Z\big)^d$. Then, you may divide this path into $N$ pieces of length $L$ each, and you have a particular state for a system that satisfies $NL = a^d$.
\\

Since a polymer is an undirected self-avoiding path, we can think of a polymer $C$ as an undirected subgraph of $\G$. However, we will sometimes need to give an orientation to the edges. This orientation is naturally induced by choosing an endpoint of the polymer. If $C = (v_1,\ldots,v_L)$, then the choice of $v_1$ will induce the orientation $v_1\rightarrow v_2 \rightarrow \ldots \rightarrow v_L$, while the choice of $v_L$ induces $v_L \rightarrow v_{L-1} \rightarrow \ldots \rightarrow v_1$.
\\

We denote by $q(\cdot)$ the targeted probability distribution on our system. Our aim is to run a Markov chain that has $q$ as stationary distribution. Typically, each state $S \in {\mathcal S}$ is assigned a probability $q(S) = Z^{-1}e^{-E(S)}$ where $E$ is an energy function, and $Z$ is the normalizing constant, also called partition function. As in the Metropolis algorithm, our probability of acceptance will depend only on ratios of values of $q$, thus making no need to know $Z$.

\subsection{Main ideas for the growth of a polymer}\label{sub:heuristic}

The \R algorithm tries to replace one polymer at each step. Hence the main difficulty that it has to address is the following: given $N-1$ polymers that do not intersect each other, how do we grow a new one? In this section, we emphasize the two salient features of the \R algorithm introduced earlier: the concepts of feeler and of underlying graph. In order to understand why these two ideas give very good results, it is helpful to have in mind the two following extreme algorithms.
\\

The first algorithm is an exhaustive algorithm, and works as follows. First, we pick a free vertex $v_1$, i.e., a vertex that is not occupied by any of the $N-1$ polymers. Then we take a free vertex $v_2$ neighbor of $v_1$ in $\G$: we have constructed a partial polymer $(v_1, v_2)$ of length $2$, and we can continue from $v_2$. Then one of the two following events happens.

Either we always have the possibility to pick up a neighboring unoccupied vertex, and we construct a polymer $(v_1, \ldots, v_L)$ which has the desired length $L$. Then we are done.

Or at some point, we have grown a partial polymer $(v_1, \ldots, v_i)$ with $1 \leq i \leq L-1$ such that $v_i$ has no free neighbor: each neighbor either belongs to some other polymer or is one of the preceding vertices $v_1, \ldots, v_{i-1}$. In this case, we recoil to $v_{i-1}$ and consider again the partial polymer $(v_1, \ldots, v_{i-1})$, but this time, we do not grow towards $v_i$. The same discussion applies again: either we have another acceptable direction to grow the partial polymer, or we have to recoil to $v_{i-2}$,~\ldots Observe that in this example, if from $v_{i-1}$ we try another direction $v'_i \neq v_i$ such that from $v'_i$ we have to recoil again, then from $v_{i-1}$, two directions would now be forbidden, namely $v_i$ and $v'_i$.

Finally, we stop either when we have grown a polymer of length $L$, in which case we successfully grew a polymer, or when we have to recoil from the very first vertex $v_1$, in which case the step was a failure. In the former case, we have randomly chosen one possible polymer starting from $v_1$, while in the latter case, there was no acceptable polymer starting from $v_1$.

This algorithm is therefore efficient in the sense that it grows successfully a polymer whenever it is possible. The chain construction rate is thus the best possible. However, due to an exhaustive search when growing the new polymer, this answer may come at the cost of very long computations, especially if $Q$ or $L$ is large, and if the system is dense.
\\

At the other extreme, the second algorithm is very fast and works identically, except that it forbids recoiling. In this algorithm, we keep picking up free neighbors at random, until either we have grown an acceptable polymer, or we have no free neighbor around the extremity of the current partial polymer. This algorithm quickly terminates, because it does at most $L-1$ trials, but the chain construction rate is very low.
\\

The concept of feeler makes it possible to tune the \R algorithm between these two extremes. Another way to look at these two algorithms is the following: for the exhaustive algorithm, no vertex on the current partial polymer $(v_1, \ldots, v_i)$, except $v_1$, is sure to be in the final polymer, if there is one. Indeed, it may happen that we recoil all the way back to $v_1$, and from $v_1$, try another direction $v'_2 \neq v_2$. For the fast algorithm, as soon as a vertex is added to the current partial polymer, it is sure to be in the final polymer, if there is one.

In the \R algorithm, a partial polymer $(v_1, \ldots, v_i)$ can always be broken up into two parts: the first part $(v_1, \ldots, v_\Delta)$ is sure to be in the potential final polymer, while the extremity $(v_{\Delta+1}, \ldots, v_i)$ serves as a retractable feeler. In other words, there is an index $\Delta$ such that the partial polymer cannot recoil below $v_\Delta$, and this index is increasing according to certain rules that we describe in details in Section~\ref{sub:rg_procedure}. The existence of a retractable feeler makes it possible to recoil from dense areas, thus improving the chain construction rate, while the existence of this index $\Delta$ keeps the computational complexity reasonable.
\\

It is essential to observe that the three algorithms described above do not depend on the graph they are run on. In particular, they can be run on any oriented graph. Indeed, all that matters is to know which directions are allowed to try to grow the polymer. However, even if these algorithms could be run on any graph, this graph  plays an important role. For instance, we have seen that the exhaustive algorithm is intractable when $Q$ is high, because the number of paths explored may be of order exponential in $Q$.
\begin{figure}[!t]
\begin{center}
	\psset{unit=1.1,linewidth=1.3pt,arrowsize=3pt 3,arrowlength=1,arrowinset=.55,arrows=->}
	\begin{pspicture}(5,5)
		\psgrid[griddots=10,gridwidth=0.8pt,subgriddiv=1,gridlabels=0]\psframe*[linecolor=white](-0.1,-0.1)(5.1,0.5)\psframe*[linecolor=white](-0.1,-0.1)(0.5,5.1)\psframe*[linecolor=white](5.1,-0.1)(4.5,5.1)\psframe*[linecolor=white](-0.1,5.1)(5.1,4.5)
		\psdot[linewidth=2pt](2,3)
		\psellipticarc{<->}(0,1)(1,0.4){0}{90}\psellipticarc{<->}(5,1)(1,0.4){90}{180}\psellipticarc(1,0)(0.4,1){90}{180}\psellipticarc(1,5)(0.4,1){180}{270}
		\psline(1,2)(1,3)\psline(1,2)(2,2)
		\psline(1,3)(1,4)\psline(1,3)(1,2)
		\psline(1,4)(1,3)\psline(1,4)(2,4)\psellipticarcn(0,4)(1,0.4){0}{270}\psellipticarcn(5,4)(1,0.4){270}{180}
		\psline(2,1)(2,2)\psline(2,1)(1,1)\psellipticarc(2,0)(0.4,1){0}{90}\psellipticarc(2,5)(0.4,1){270}{0}
		\psline(2,2)(2,3)\psline(2,2)(1,2)
		\psline(2,3)(2,4)\psline(2,3)(3,3)
		\psline(2,4)(1,4)
		\psline(3,2)(3,3)\psline(3,2)(2,2)
		\psline(3,3)(3,2)\psline(3,3)(3,4)
		\psline(3,4)(2,4)\psline(3,4)(4,4)
		\psline(4,4)(3,4)\psellipticarcn(4,0)(0.4,1){90}{0}\psellipticarcn(4,5)(0.4,1){0}{270}
	\end{pspicture}
\end{center}
	\vspace{-0.5cm}
	\caption{Example of an underlying graph in dimension $2$ with parameter $k=2$. The dashed lines represent the underlying lattice. We can see that each vertex has exactly $k=2$ out-edges. The dotted vertex is the root.}\label{fig:example_underlying}
\end{figure}
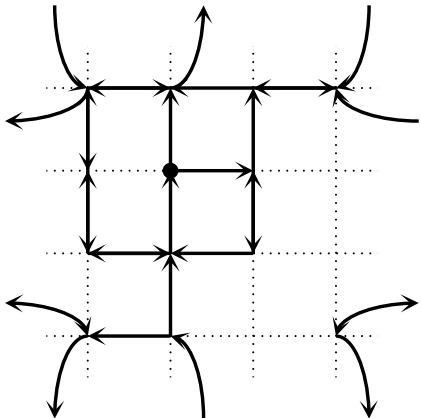
But since the algorithm that generates a polymer may be run on any graph, it is therefore a very natural idea to run it on some oriented subgraph $G$ instead of $\G$ itself. The graph on which the algorithm is actually run is precisely what we call the underlying graph. Since we consider subgraphs of $\G$, the out-degree of each vertex in an underlying graph is at most $Q$. The simplest  underlying graphs, and the ones that we consider until Section~\ref{subsub:random}, are graphs that have a constant out-degree $k$, meaning that each vertex of the subgraph has the same out-degree $k \leq Q$. For $k < Q$, the sets of allowable directions are actually reduced, leading to a gain in complexity. However, by forbidding some directions that could potentially lead to an admissible polymer, the construction rate is reduced as well.

Figure~\ref{fig:example_underlying} shows an example of an underlying graph with parameter $k=2$ in dimension $d=2$. We see that each vertex has indeed exactly two out-edges, and some edges are curved to highlight the torus structure of the underlying lattice. Finally, one can see a dot on a vertex: this vertex is the root of the underlying graph. It is a special vertex that indicates where to begin the growth of the polymer. More details on the structure and generation of underlying graphs are given in Section~\ref{sub:underlying}.

Similarly as for $\ell$, the length of the feeler, the higher $k$, the higher the construction rate and the higher the computational cost. So here again, a trade-off has to be made.

\section{The Recoil Growth Algorithm to generate multi-polymer systems}\label{sec:rga}

\subsection{Overview}\label{sub:overview}

With the heuristic description made in Section~\ref{sub:heuristic}, we can now give an overview of the \R algorithm. At the beginning of the algorithm, the $N$ polymers are placed on $\G$, for instance as the result of the procedure given in Section~\ref{sub:notations}, but any other state is fine.

Assume that the system is currently in the state $S_\old$, defined by the configurations of the $N$ different polymers on the underlying lattice $\G$. The first step of the algorithm is to choose a polymer $C_\old$ uniformly at random among the $N$ possible, and to remove it from the system. In this modified system, with only $N-1$ polymers left, we now try to grow a new polymer $C_\new$. In order to do so, we first generate at random an underlying graph $G_\new$, then we try to grow a new polymer $C_\new$ on $G_\new$: this is the \R procedure. At this point, two things may happen:
\\

\noindent (i) the growth is not successful: then this step is a failure, and we go back to the original state $S_\old$ by putting back $C_\old$.
\\

\noindent (ii) the growth is successful: we now have two states $S_\old$ and $S_\new$ that only differ by one polymer, $C_\old$ in $S_\old$ and $C_\new$ in $S_\new$. According to the Metropolis algorithm, the right stationary distribution $q$ is obtained by accepting the new state $S_\new$ with a suitable probability. But here, by conditioning the generation of $C_\new$ according to a graph $G_\new$, we introduced a skewness between $S_\old$ and $S_\new$. By analogy with the auxiliary variable method, we need to generate an underlying graph $G_\old$ that plays for $S_\old$ the role that $G_\new$ plays for $S_\new$. This generation is the next step of the \R algorithm. Then the right probability of acceptance is not the probability of accepting $S_\new$ compared to $S_\old$, but the probability to accept $(S_\new, G_\new)$ compared to $(S_\old, G_\old)$ that we denote by $P_\acc\big((S_\old, G_\old), (S_\new, G_\new)\big)$. To compute this probability, we first need to compute the weights of $C_\old$ and $C_\new$ on $G_\old$ and $G_\new$ respectively. Weights are complex objects that are defined in Section~\ref{sub:rg_procedure}. Because they are directly connected to the probability $P_\RG(C|G)$ of generating the polymer $C$ on the underlying graph $G$, they are nonetheless necessary in order to compute the right probability of acceptance.

For the sake of notations, in the remainder of this paper, the subscripts $\old$ and $\new$ will be noted $\o$ and $\n$ respectively, so for instance $C_\o$ denotes $C_\old$.

One step of the \R algorithm can be summarized as follows:

\noindent\rule{\textwidth}{0.3mm}
\vspace{-0.4cm}
\texttt{
	\begin{enumerate}[leftmargin=*, rightmargin=1cm]
		\item Choose a polymer $C_\o$ w.p.\ $1/N$ and remove it from the system
		\item Generate at random an underlying graph $G_\n$ w.p.\ $P_\U(G_\n)$
	\item Run the \R procedure on $G_\n$
	\begin{enumerate}
		\item If the procedure did not successfully grow a new polymer:
		\begin{enumerate}
			\item Put back $C_\o$
			\item Go to 1.
		\end{enumerate}
		\item Else, we have grown a new polymer $C_\n$ w.p.\ $P_\RG(C_\n|G_\n)$:
		\begin{enumerate}
			\item Generate at random an underlying graph $G_\o$ compatible with $C_\o$ w.p.\ $P_\C(G_\o|C_\o)$
			\item Compute the weights $W(C_\n|G_\n)$ and $W(C_\o|G_\o)$
			\item Set $p = P_\acc\big((S_\o, G_\o), (S_\n, G_\n)\big)$, flip a $p$-coin:
			\begin{enumerate}
				\item If heads, add $C_\n$ to the polymer system
				\item If tails, put back $C_\o$
			\end{enumerate}
			\item Go to 1.
			\end{enumerate}
		\end{enumerate}
	\end{enumerate}
}
\vspace{-0.2cm}
\noindent\rule{\textwidth}{0.3mm}
\\

With this description of our algorithm, it is straightforward to write down the transition probability $P(S_\o \rightarrow S_\n)$ from $S_\o$ to $S_\n$ where $S_\o$ and $S_\n$ differ by only one polymer $C_\o$ in $S_\o$ and $C_\n$ in $S_\n$:
\begin{align}\label{eq:transition}
  P(S_\o \rightarrow S_\n) = \frac{1}{N}\sum_{G_\n,\ G_\o}P_\U(G_\n) P_\RG(C_\n|G_\n)P_\C(G_\o|C_\o) P_\acc\big((S_\o, G_\o), (S_\n, G_\n)\big).
\end{align}
Each term of this sum will be computed below.

A complete description of the \R algorithm, to be given next, is as follows: (i) define precisely the concept of underlying graph as well as the notion of compatibility between an underlying graph and a polymer; (ii) explain the \R procedure, in which the concept of feeler plays a central role, and finally (iii) define the weight of a polymer on an underlying graph, which makes it possible to to compute the right probability of acceptance.

\subsection{Underlying graphs}\label{sub:underlying}

\begin{definition}
  An underlying graph $G \subset \G$ is a directed subgraph of $\G$ rooted at a vertex $r$ such that:
  \begin{enumerate}
    \item for any vertex $v \in G\backslash\{r\}$, there exists a directed path from $r$ to $v$
    \item each vertex has exactly $k$ out-edges
  \end{enumerate}
\end{definition}

\noindent\textbf{Remark:} Several vertices may satisfy the same property as the root, namely accessibility to all other vertices. However, the root plays a special role, as it will be the starting vertex for the growth of a polymer.
\\

As mentioned earlier, $1 \leq k \leq Q$ is a parameter of the algorithm. Note that when $k = Q$, every underlying graph spans the whole underlying lattice with every possible directed edge induced by $\G$. So in this case, an underlying graph is simply characterized by its root. When $k=1$, an underlying graph is just an oriented path starting at the root, and ending as soon as a loop appears. Figure~\ref{fig:example_underlying} shows an example of underlying graph with $k=2$ in dimension 2. The dotted vertex is the root, and one can check that there exists a path from the root to any vertex of the underlying graph.

\begin{definition}
  We say that a polymer $C$ and an underlying graph $G$ are \textnormal{compatible} if the root $r$ of $G$ is one of the two extremities of $C$ and if $C \subset G$ (where the choice of $r$ induces the orientation on $C$ as explained in Section~\ref{sub:notations}).
\end{definition}

The class of underlying graphs is the natural class of graphs to run the \R procedure; as we will see, a polymer can be grown on some underlying graph if and only if they are compatible.

\subsubsection{Procedure to generate an underlying graph}\label{subsub:generation_underlying}

After choosing a random polymer to be removed from the system, the first step of the \R algorithm is to generate an underlying graph. The growth of the candidate polymer will be performed on this underlying graph.

To generate an underlying graph, we proceed iteratively, taking care that every vertex added to the graph has exactly $k$ out-edges. This algorithm is basically a greedy algorithm that at each step assigns out-edges to some randomly chosen vertex.
\\

More precisely, we have two sets $V$ and $T$: $V$ is the set of vertices of the final underlying graph that we are trying to generate, whereas $T$ is a waiting room for the vertices that will eventually end up in $V$, but have not been assigned out-edges yet.

Initially, we choose the root $r$ uniformly at random in the set of vertices that are not occupied by any of the $N-1$ polymers in the system. Thus $T = \{r\}$ and $V = \emptyset$ because we have not assigned out-edges to the root yet. The first step is to assign $k$ out-edges to $r$: $r$ has $Q$ neighbors in $\G$, and we choose $k$ out-edges out of these $Q$ possible uniformly at random. Call $v_1,\ldots, v_k$ the corresponding neighbors or $r$. Then we have assigned out-edges to $r$ but not to the $k$ new vertices, so $T = \{v_1,\ldots, v_k\}$ and $V = \{r\}$. Then we take any vertex in $T$, say $v_1$, and we choose its $k$ neighbors. Since $r$ and $v_1$ are neighbors in $\G$, it may happen that $r$ is among these neighbors, so we have to be careful: we add to $T$ the vertices that are not already in $V$ or in $T$, and we move $v_1$ from $T$ to $V$. We keep on assigning out-edges to vertices in $T$ until $T = \emptyset$.

We see that in this algorithm, except for the root, the current configuration of the system is not taken into account. We add vertices and out-edges independently of the polymers on $\G$. This leads to some inefficiency, and we explain in Section~\ref{sec:implementation} how to remedy to this problem in practice.
\\

When $T$ is empty, it is clear that each vertex in $V$ has exactly $k$ out-edges, and that there exists a directed path from $r$ to any vertex: we have indeed generated an underlying graph. Observe that there is no constraint on the in-degree of each vertex. However, since there is a path from the root to any vertex but the root, each vertex except the root has in-degree at least 1. The root is the only vertex that may have in-degree 0.

In the following pseudo-code that explains how to generate an underlying graph, $E$ is the set of edges we are trying to generate, and $V$ and $T$ are as in the above informal description:

\noindent\rule{\textwidth}{0.3mm}
\vspace{-0.4cm}
\texttt{
	\begin{enumerate}[leftmargin=*, rightmargin=0.8cm]
		\item Pick a free vertex $r$ at random, and initialize as follows:
		\begin{enumerate}
			\item $V = E = \emptyset$
			\item $T = \{r\}$
		\end{enumerate}
		\item While $T \neq \emptyset$:
		\begin{enumerate}
			\item Pick any vertex $v$ in $T$
			\item Choose $k$ different neighbors $(v_1,\ldots,v_k)$ of $v$ uniformly at random
			\item For $i = 1 \ldots k$
			\begin{enumerate}
				\item Add the edge $v \rightarrow v_i$ to $E$
				\item If $v_i \notin V$ and $v_i \notin T$, add $v_i$ to $T$
			\end{enumerate}
			\item Remove $v$ from $T$ and add it to $V$
		\end{enumerate}
	\end{enumerate}
}
\vspace{-0.2cm}
\noindent\rule{\textwidth}{0.3mm}
\\

\begin{figure}
	\centering
	\psset{unit=0.8,linewidth=1.3pt,arrowsize=3pt 3,arrowlength=1,arrowinset=.55}
	\subfloat[][]{\label{subfig:1}
		\begin{pspicture}(5,5)
			\psgrid[griddots=10,gridwidth=0.8pt,subgriddiv=1,gridlabels=0]\psframe*[linecolor=white](-0.1,-0.1)(5.1,0.5)\psframe*[linecolor=white](-0.1,-0.1)(0.5,5.1)\psframe*[linecolor=white](5.1,-0.1)(4.5,5.1)\psframe*[linecolor=white](-0.1,5.1)(5.1,4.5)
			\psdot[linewidth=2pt](2,3)
			\rput(2,3){\psline(-0.2,-0.2)(0.2,0.2)\psline(-0.2,0.2)(0.2,-0.2)}

			\psline(1,3)(1,2)\psellipticarc(0,2)(1,0.3){0}{90}\psellipticarc(5,2)(1,0.3){90}{180}\psline(4,2)(3,2)(3,1)
			\psline(4,3)(3,3)(3,4)(4,4)\psellipticarc(4,5)(0.3,1){180}{270}\psellipticarc(4,0)(0.3,1){90}{180}
		\end{pspicture}
	}
	\subfloat[][]{
		\begin{pspicture}(5,5)
			\psgrid[griddots=10,gridwidth=0.8pt,subgriddiv=1,gridlabels=0]\psframe*[linecolor=white](-0.1,-0.1)(5.1,0.5)\psframe*[linecolor=white](-0.1,-0.1)(0.5,5.1)\psframe*[linecolor=white](5.1,-0.1)(4.5,5.1)\psframe*[linecolor=white](-0.1,5.1)(5.1,4.5)
			\psdot[linewidth=2pt](2,3)
			\psline{->}(2,3)(2,4)\psline{->}(2,3)(3,3)
			\rput(2,4){\psline(-0.2,-0.2)(0.2,0.2)\psline(-0.2,0.2)(0.2,-0.2)}
			\rput(3,3){\psline(-0.2,-0.2)(0.2,0.2)\psline(-0.2,0.2)(0.2,-0.2)}
		\end{pspicture}
	}
	\\
	\subfloat[][]{
		\begin{pspicture}(5,5)
			\psgrid[griddots=10,gridwidth=0.8pt,subgriddiv=1,gridlabels=0]\psframe*[linecolor=white](-0.1,-0.1)(5.1,0.5)\psframe*[linecolor=white](-0.1,-0.1)(0.5,5.1)\psframe*[linecolor=white](5.1,-0.1)(4.5,5.1)\psframe*[linecolor=white](-0.1,5.1)(5.1,4.5)
			\psdot[linewidth=2pt](2,3)
			\rput(1,4){\psline(-0.2,-0.2)(0.2,0.2)\psline(-0.2,0.2)(0.2,-0.2)}
			\rput(2,1){\psline(-0.2,-0.2)(0.2,0.2)\psline(-0.2,0.2)(0.2,-0.2)}
			\psline{->}(2,3)(2,4)\psline{->}(2,3)(3,3)
			\psline{->}(2,4)(1,4)\psellipticarc{->}(2,5)(0.3,1){270}{0}\psellipticarc{->}(2,0)(0.3,1){0}{90}
			\rput(3,3){\psline(-0.2,-0.2)(0.2,0.2)\psline(-0.2,0.2)(0.2,-0.2)}
		\end{pspicture}
	}
	\subfloat[][]{
		\begin{pspicture}(5,5)
			\psgrid[griddots=10,gridwidth=0.8pt,subgriddiv=1,gridlabels=0]\psframe*[linecolor=white](-0.1,-0.1)(5.1,0.5)\psframe*[linecolor=white](-0.1,-0.1)(0.5,5.1)\psframe*[linecolor=white](5.1,-0.1)(4.5,5.1)\psframe*[linecolor=white](-0.1,5.1)(5.1,4.5)
			\psdot[linewidth=2pt](2,3)
			\rput(1,1){\psline(-0.2,-0.2)(0.2,0.2)\psline(-0.2,0.2)(0.2,-0.2)}
			\rput(1,4){\psline(-0.2,-0.2)(0.2,0.2)\psline(-0.2,0.2)(0.2,-0.2)}
			\psline{->}(2,1)(2,2)\psline{->}(2,1)(1,1)
			\rput(2,2){\psline(-0.2,-0.2)(0.2,0.2)\psline(-0.2,0.2)(0.2,-0.2)}
			\psline{->}(2,3)(2,4)\psline{->}(2,3)(3,3)
			\psline{->}(2,4)(1,4)\psellipticarc{->}(2,5)(0.3,1){270}{0}\psellipticarc{->}(2,0)(0.3,1){0}{90}
			\rput(3,3){\psline(-0.2,-0.2)(0.2,0.2)\psline(-0.2,0.2)(0.2,-0.2)}
		\end{pspicture}
	}
	\\
	\subfloat[][]{
		\begin{pspicture}(5,5)
			\psgrid[griddots=10,gridwidth=0.8pt,subgriddiv=1,gridlabels=0]\psframe*[linecolor=white](-0.1,-0.1)(5.1,0.5)\psframe*[linecolor=white](-0.1,-0.1)(0.5,5.1)\psframe*[linecolor=white](5.1,-0.1)(4.5,5.1)\psframe*[linecolor=white](-0.1,5.1)(5.1,4.5)
			\psdot[linewidth=2pt](2,3)
			\psellipticarc{->}(0,1)(1,0.3){0}{90}\psellipticarc{->}(5,1)(1,0.3){90}{180}\psellipticarc{->}(1,0)(0.3,1){90}{180}\psellipticarc{->}(1,5)(0.3,1){180}{270}
			\rput(1,4){\psline(-0.2,-0.2)(0.2,0.2)\psline(-0.2,0.2)(0.2,-0.2)}
			\psline{->}(2,1)(2,2)\psline{->}(2,1)(1,1)
			\rput(2,2){\psline(-0.2,-0.2)(0.2,0.2)\psline(-0.2,0.2)(0.2,-0.2)}
			\psline{->}(2,3)(2,4)\psline{->}(2,3)(3,3)
			\psline{->}(2,4)(1,4)\psellipticarc{->}(2,5)(0.3,1){270}{0}\psellipticarc{->}(2,0)(0.3,1){0}{90}
			\rput(3,3){\psline(-0.2,-0.2)(0.2,0.2)\psline(-0.2,0.2)(0.2,-0.2)}
			\rput(4,1){\psline(-0.2,-0.2)(0.2,0.2)\psline(-0.2,0.2)(0.2,-0.2)}
		\end{pspicture}
	}
	\subfloat[][]{\label{subfig:last}
		\begin{pspicture}(5,5)
			\psgrid[griddots=10,gridwidth=0.8pt,subgriddiv=1,gridlabels=0]\psframe*[linecolor=white](-0.1,-0.1)(5.1,0.5)\psframe*[linecolor=white](-0.1,-0.1)(0.5,5.1)\psframe*[linecolor=white](5.1,-0.1)(4.5,5.1)\psframe*[linecolor=white](-0.1,5.1)(5.1,4.5)
			\psdot[linewidth=2pt](2,3)
			\psellipticarc{->}(0,1)(1,0.3){0}{90}\psellipticarc{->}(5,1)(1,0.3){90}{180}\psellipticarc{->}(1,0)(0.3,1){90}{180}\psellipticarc{->}(1,5)(0.3,1){180}{270}
			\rput(1,2){\psline(-0.2,-0.2)(0.2,0.2)\psline(-0.2,0.2)(0.2,-0.2)}
			\rput(1,4){\psline(-0.2,-0.2)(0.2,0.2)\psline(-0.2,0.2)(0.2,-0.2)}
			\psline{->}(2,1)(2,2)\psline{->}(2,1)(1,1)
			\psline{->}(2,2)(2,3)\psline{->}(2,2)(1,2)
			\psline{->}(2,3)(2,4)\psline{->}(2,3)(3,3)
			\psline{->}(2,4)(1,4)\psellipticarc{->}(2,5)(0.3,1){270}{0}\psellipticarc{->}(2,0)(0.3,1){0}{90}
			\rput(3,3){\psline(-0.2,-0.2)(0.2,0.2)\psline(-0.2,0.2)(0.2,-0.2)}
			\rput(4,1){\psline(-0.2,-0.2)(0.2,0.2)\psline(-0.2,0.2)(0.2,-0.2)}
		\end{pspicture}
	}
	\caption{The six first steps of the generation of an underlying graph. \protect\subref{subfig:1}: the root $r$ is the dotted vertex, which we pick uniformly at random among the unoccupied vertices. The edges represent the $N-1 = 2$ polymers left in the system. In the sequel of the generation of the underlying graph, the polymers play no role, so they are not drawn. \protect\subref{subfig:1}-\protect\subref{subfig:last}: vertices with a cross will be in the final underlying graph, but have not been assigned out-edges to yet: they form the set $T$. At each step, we pick one vertex in $T$ at random, and assign out-edges to it. The vertex then becomes part of the set $V$, and we remove the cross. If by assigning out-edges, we have discovered new vertices, we add these vertices to $T$. In~\protect\subref{subfig:last}, the generation is not over, since $T$ has four more vertices.}\label{fig:underlying-graph}
\end{figure}
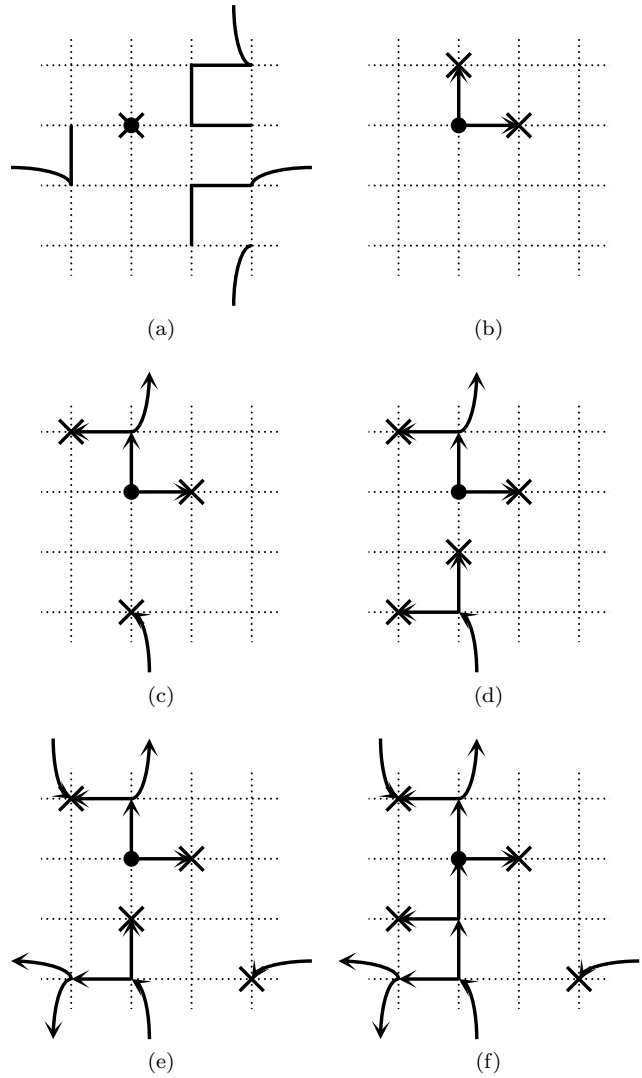

Since the underlying lattice $\G$ is finite, it is clear that this procedure terminates in finite time. Moreover, it is easy to see that this procedure does not depend on the order in which you pick up vertices in $T$: once a vertex is in $T$, its set of out-edges does not depend on when you assign them. Figure~\ref{fig:underlying-graph} page~\pageref{fig:underlying-graph} shows the six first steps of the generation of an underlying graph in dimension two with $k=2$.

\begin{proposition}\label{prop:P_U}
  The probability $P_\U(G)$ to generate the underlying graph $G$ is given by
  \begin{align}\label{eq:P_U}
    P_\U(G) = \frac{\alpha^{|G|}}{\gamma}
  \end{align}
  where $\alpha = \binom{Q}{k}^{-1}$, $|G|$ is the number of vertices in $G$, and $\gamma = a^d-(N-1)L$ is the number of free vertices when $N-1$ polymers are in the system.
\end{proposition}

\begin{proof}
The generation of $G$ is unambiguous: the root is chosen with probability $1/\gamma$; then, each time we assign out-edges, we have to assign the right set of out-edges. Since there are $1/\alpha$ such sets and the choice is uniform random, the probability of assigning the right set of out-edges is exactly $\alpha$. Such a choice occurs for each of the $|G|$ vertices, hence the result.
\end{proof}

\subsubsection{Procedure to generate a compatible underlying graph}

As we have seen in the overview of the \R algorithm in Section~\ref{sub:overview}, the transition between the old and the new state is made symmetric by generating an underlying graph compatible with the old polymer $C_\o$, as a comparison to the graph $G_\n$.

The procedure to do so is essentially the same as to generate any underlying graph: the only difference is that when we try to assign out-edges to vertices that belong to $C_\o = (v_1, \ldots, v_L)$, we must take care that for each $1 \leq i \leq L-1$, we do have $v_i \rightarrow v_{i+1}$ in the set of out-edges. Hence for these $L-1$ vertices, one out-edge is imposed, and we have to choose the $k-1$ others among $Q-1$ possible. Hence we can derive the following property similarly as for general underlying graphs:

\begin{proposition}
  Given a polymer $C$ of length $L$, the probability $P_\C(G|C)$ to generate an underlying graph $G$ compatible with $C$ is given by
  \begin{align}\label{eq:P_C}
    P_\C(G|C) = \frac{\alpha^{|G|- L + 1} \beta^{L-1}}{2}
  \end{align}
  where $\beta = \binom{Q-1}{k-1}^{-1}$ and $\alpha$ and $|G|$ are as in Property~\ref{prop:P_U}.
\end{proposition}

\noindent\textbf{Remark:} $P_\C(G|C)$ and $P_\U(G)$ are proportional: we have $P_\C(G|C) = \eta P_\U(G)$ where $\eta = \gamma/2 \cdot (\beta/\alpha)^{L-1}$ is a universal constant that does not depend on the polymer or the underlying graph.

\subsection{Recoil Growth Procedure}\label{sub:rg_procedure}

In this section, we describe in details the procedure that grows a polymer on a given underlying graph $G$. As explained before, this growth takes place on an underlying graph that has a constant out-degree $k$. Before describing the dynamics according to which the algorithm either extends an existing partial polymer or recoils, we have to understand the concept of feeler.

\subsubsection{Decomposition of a partial polymer $\overline C$}

During the growing process, a partial polymer $\overline C = (v_1, \ldots, v_i)$ of length $i < L$ can always be decomposed into two disjoint parts: the \emph{fixed part} and the \emph{feeler}. Before stating the definition of this decomposition, it is important to have in mind the rule that this decomposition stands for:
\\

\noindent\textbf{Rule:} \textit{During the growth process, we are allowed to recoil from a vertex if and only if it is in the feeler.}
\\

\noindent This means that if at some point we need to recoil from a vertex that belongs to the fixed part, the attempt is a failure. Since the feeler is the complementary part of the fixed part, one only needs to define the fixed part:

\begin{definition}
	A node $v_j$ is in the fixed part if $j=1$ or if during the history of the growth of $\overline C$, there was a partial polymer  $(v_1, \ldots, v_j, v_{j+1}, \ldots, v_{j+\ell})$.
\end{definition}
Equivalently, the fixed part is the set of vertices $v$ such that either $v$ is the starting vertex of the polymer, or at some point in the growth history, a partial polymer has been grown through $v$ and has been $\ell$ steps further. The previous definition immediately yields the following properties:

\begin{proposition}
	There exists an increasing index $\Delta$ such that the fixed part is of the form $(v_1, \ldots, v_\Delta)$.
\end{proposition}

\begin{proof}
Consider a vertex $v_j$ in the fixed part, and the corresponding partial polymer $\overline C = (v_1, \ldots, v_j, v_{j+1}, \ldots, v_{j+\ell})$. Since the \R procedure extends or shortens a partial polymer only one vertex at a time, this implies that necessarily, at some point, the shorter polymer $(v_1, \ldots,v_{j-1}, v_j, \ldots, v_{j+\ell-1})$ was grown: this exactly means that $j-1$ is in the fixed part as well. This proves that the fixed part is indeed of the form $(v_1, \ldots, v_\Delta)$.

Moreover, it is clear by definition that if a vertex belongs at some point to the fixed part, then it stays in the fixed part for the remainder of the growth process: the fixed part is increasing, hence so is the index $\Delta$.
\end{proof}

\begin{proposition}
	The feeler is of the form $(v_{\Delta+1}, \ldots, v_i)$, and is of length at most $\ell$ (and is possibly empty).
\end{proposition}

\begin{proof}
By the previous proposition, all we have to show is that the feeler is indeed of length at most $\ell$. Imagine for a moment that the feeler is of length $\ell+1$: this means that we have a partial polymer $\overline C = (v_1, \ldots, v_\Delta, v_{\Delta+1}, v_{\Delta+2}, \ldots, v_{\Delta+\ell+1})$ where $\Delta$ delimits the fixed part. But then, by definition of the fixed part, the vertex $\Delta+1$ should belong to the fixed part, which is not possible.
\end{proof}

We can now draw a picture of the way the fixed part and the feeler evolve. If $\overline C = (v_1, \ldots, v_i)$ grows towards $v_{i+1}$, the feeler is increased except when it was already of length $\ell$. If the feeler is of length $\ell$ when $\overline C$ grows, then it stays of length $\ell$, and the fixed part grows. It is then straightforward to derive the formula $\Delta = \max(1, L_{\mathrm{max}} - \ell)$, where $L_{\mathrm{max}}$ is the length of the longest polymer grown in the history of $C$.

Note that vertices $v_j$ with $j \geq L-\ell$ are never in the fixed part, because we never grow a polymer of length larger than $L$.
\\

By playing on $\ell$, we can have the two extreme algorithms described in Section~\ref{sub:heuristic}. For $\ell = 0$, we get the fast algorithm, because since the feeler is of length $0$, there is no vertex from which it is allowed to recoil. And if one sets $\ell = L$, it is always allowed to recoil from any vertex, and we get the exhaustive algorithm. So this parameter is the parameter that makes it possible to make a trade-off between the speed and the tractability of the algorithm.

We can now fairly simply describe the \R procedure given an underlying graph~$G$.

\subsubsection{Description of the \R procedure}

Given an underlying graph $G$, the \R procedure keeps record of the current partial polymer $\overline C$, the length $L_{\mathrm{max}}$ of the longest partial polymer ever grown, and $L$ sets $D_i$. For a partial polymer of length at least $i$, $D_i$ represents the set of vertices that the vertex $v_i$ has already tried as directions of growth. In particular, at any time, $D_i$ is a subset of the set of neighbors of $v_i$ in $G$. These sets are here to insure that the algorithm is free of loop: when you recoil to a vertex $v_i$, you must try a new direction, that you have not tried so far. If at some point, you recoil to $v_i$ and every neighbor of $v_i$ is in $D_i$ (equivalently, $|D_i| = k$), this means that you have again to recoil from $v_i$. This is possible only when $v_i$ is in the feeler, which is easily verified by checking the condition $i > \max(1, L_{\mathrm{max}}-\ell)$.

The \R procedure works as follows:

\noindent\rule{\textwidth}{0.3mm}
\vspace{-0.4cm}
\texttt{
	\begin{enumerate}[leftmargin=*, rightmargin=0.8cm]
		\item Initialization:
		\begin{enumerate}
			\item $\overline C = (v_1)$, where $v_1$ is the root of $G$
			\item $D_1 = \emptyset$
			\item $L_{\mathrm{max}} = 1$
		\end{enumerate}
		\item At each step, with $\overline C = (v_1, \ldots, v_i)$:
		\begin{enumerate}
			\item If $i = L$, stop, $\overline C$ is a complete polymer
			\item Else if $|D_i| = k$, recoil:
			\begin{enumerate}
				\item If $i > \max(1, L_{\mathrm{max}}-\ell)$, set $\overline C = (v_1, \ldots, v_{i-1})$ and go to~2
				\item Otherwise, stop, the generation has failed
			\end{enumerate}
			\item Else, keep picking uniformly at random a vertex $v$ neighbor of $v_i$ in $G$ and not in $D_i$, add it to $D_i$, and stop when either $v$ is unoccupied or $|D_i| = k$: 
			\begin{enumerate}
				\item If $v$ is unoccupied, set $\overline C = (v_1, \ldots, v_i, v)$, $D_{i+1} = \emptyset$, update $L_{\mathrm{max}}$ and go to 2
				\item Else, $|D_i| = k$, and recoil as specified in 2.(b)
			\end{enumerate}
		\end{enumerate}
	\end{enumerate}
}
\vspace{-0.2cm}
\noindent\rule{\textwidth}{0.3mm}
\\

It is not hard to show that this procedure is free of loop, and therefore terminates in finite time. The absence of loops is insured by the sets $D_i$: you cannot grow twice the same polymer because this would imply to choose as next direction a vertex that already belongs to some set $D_i$.

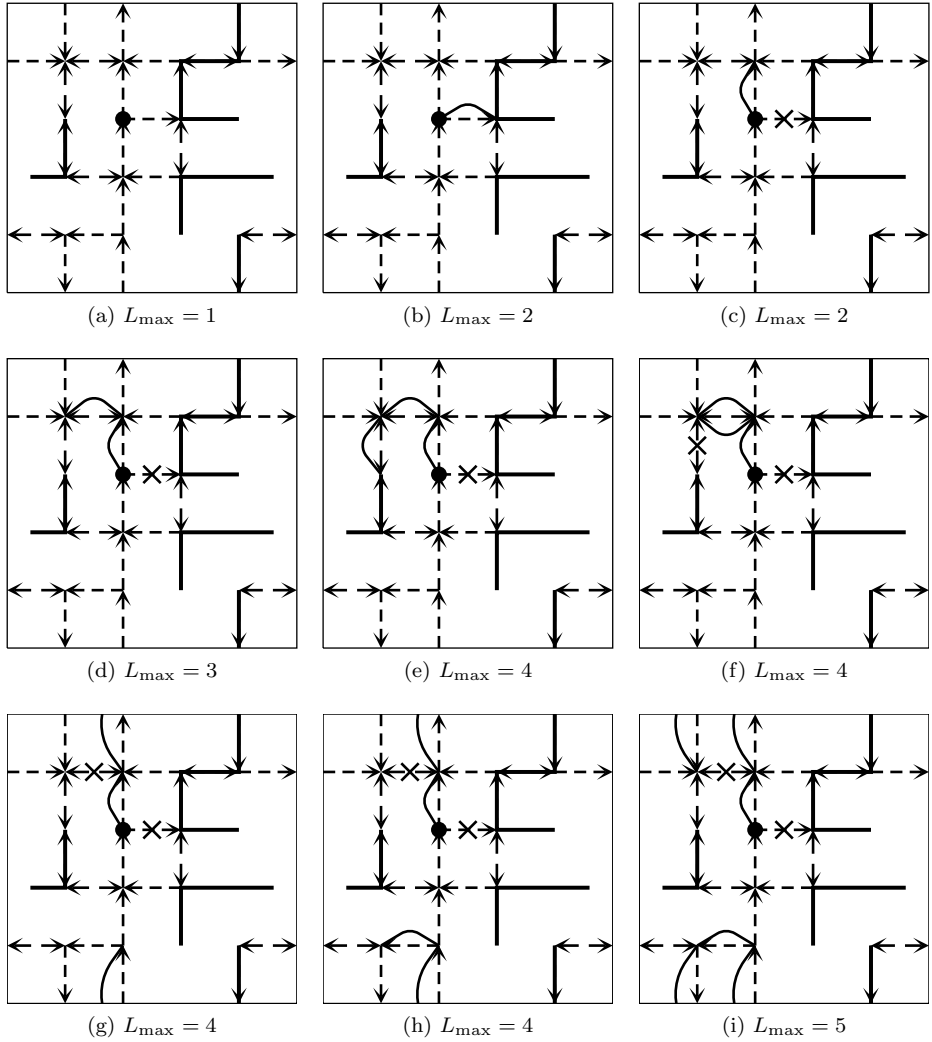
\begin{figure}
	\centering
	\psset{unit=0.77,linewidth=1pt,arrowsize=3pt 3,arrowlength=1,arrowinset=.55}
	\newpsstyle{latticestyle}{arrows=->,linestyle=dashed}
	\newpsstyle{polymerstyle}{linewidth=1.5pt}
	\newpsstyle{growthstyle}{linearc=0.3}

	\subfloat[][$L_{\mathrm{max}}=1$]{\label{subfig:a}
		\begin{pspicture}(5,5)
			\psline[linewidth=0.5pt](0,0)(5,0)(5,5)(0,5)(0,0)

			\psdot[linewidth=2pt](2,3)
			\psline[style=latticestyle](1,1)(1,0)\psline[style=latticestyle]{<->}(1,1)(0,1)\psline[style=latticestyle]{<->}(5,1)(4,1)\psline[style=latticestyle](1,5)(1,4)
			\psline[style=latticestyle]{<->}(1,2)(1,3)\psline[style=latticestyle]{<->}(1,2)(2,2)
			\psline[style=latticestyle]{<->}(1,4)(1,3)
			\psline[style=latticestyle](2,1)(2,2)\psline[style=latticestyle](2,1)(1,1)
			\psline[style=latticestyle](2,2)(2,3)
			\psline[style=latticestyle](2,3)(2,4)\psline[style=latticestyle](2,3)(3,3)
			\psline[style=latticestyle](2,4)(2,5)\psline[style=latticestyle]{<->}(2,4)(1,4)\psline[style=latticestyle](2,0)(2,1)
			\psline[style=latticestyle]{<->}(3,2)(3,3)\psline[style=latticestyle](3,2)(2,2)
			\psline[style=latticestyle](3,3)(3,4)
			\psline[style=latticestyle](3,4)(2,4)
			\psline[style=latticestyle](4,1)(4,0)\psline[style=latticestyle](4,5)(4,4)
			\psline[style=latticestyle]{<->}(4,4)(3,4)\psline[style=latticestyle](4,4)(5,4)\psline[style=latticestyle](0,4)(1,4)

			\psline[style=polymerstyle](1,3)(1,2)(0.4,2)\psline[style=polymerstyle](4.6,2)(4,2)(3,2)(3,1)
			\psline[style=polymerstyle](4,3)(3,3)(3,4)(4,4)(4,5)\psline[style=polymerstyle](4,0)(4,1)
		\end{pspicture}
	}
	\subfloat[][$L_{\mathrm{max}}=2$]{\label{subfig:b}
		\begin{pspicture}(5,5)
			\psline[linewidth=0.5pt](0,0)(5,0)(5,5)(0,5)(0,0)

			\psdot[linewidth=2pt](2,3)
			\psline[style=latticestyle](1,1)(1,0)\psline[style=latticestyle]{<->}(1,1)(0,1)\psline[style=latticestyle]{<->}(5,1)(4,1)\psline[style=latticestyle](1,5)(1,4)
			\psline[style=latticestyle]{<->}(1,2)(1,3)\psline[style=latticestyle]{<->}(1,2)(2,2)
			\psline[style=latticestyle]{<->}(1,4)(1,3)
			\psline[style=latticestyle](2,1)(2,2)\psline[style=latticestyle](2,1)(1,1)
			\psline[style=latticestyle](2,2)(2,3)
			\psline[style=latticestyle](2,3)(2,4)\psline[style=latticestyle](2,3)(3,3)
			\psline[style=latticestyle](2,4)(2,5)\psline[style=latticestyle]{<->}(2,4)(1,4)\psline[style=latticestyle](2,0)(2,1)
			\psline[style=latticestyle]{<->}(3,2)(3,3)\psline[style=latticestyle](3,2)(2,2)
			\psline[style=latticestyle](3,3)(3,4)
			\psline[style=latticestyle](3,4)(2,4)
			\psline[style=latticestyle](4,1)(4,0)\psline[style=latticestyle](4,5)(4,4)
			\psline[style=latticestyle]{<->}(4,4)(3,4)\psline[style=latticestyle](4,4)(5,4)\psline[style=latticestyle](0,4)(1,4)

			\psline[style=polymerstyle](1,3)(1,2)(0.4,2)\psline[style=polymerstyle](4.6,2)(4,2)(3,2)(3,1)
			\psline[style=polymerstyle](4,3)(3,3)(3,4)(4,4)(4,5)\psline[style=polymerstyle](4,0)(4,1)

			\psline[style=growthstyle](2,3)(2.5,3.3)(3,3)
		\end{pspicture}
	}
	\subfloat[][$L_{\mathrm{max}}=2$]{\label{subfig:c}
		\begin{pspicture}(5,5)
			\psline[linewidth=0.5pt](0,0)(5,0)(5,5)(0,5)(0,0)

			\psdot[linewidth=2pt](2,3)
			\psline[style=latticestyle](1,1)(1,0)\psline[style=latticestyle]{<->}(1,1)(0,1)\psline[style=latticestyle]{<->}(5,1)(4,1)\psline[style=latticestyle](1,5)(1,4)
			\psline[style=latticestyle]{<->}(1,2)(1,3)\psline[style=latticestyle]{<->}(1,2)(2,2)
			\psline[style=latticestyle]{<->}(1,4)(1,3)
			\psline[style=latticestyle](2,1)(2,2)\psline[style=latticestyle](2,1)(1,1)
			\psline[style=latticestyle](2,2)(2,3)
			\psline[style=latticestyle](2,3)(2,4)\psline[style=latticestyle](2,3)(3,3)
			\psline[style=latticestyle](2,4)(2,5)\psline[style=latticestyle]{<->}(2,4)(1,4)\psline[style=latticestyle](2,0)(2,1)
			\psline[style=latticestyle]{<->}(3,2)(3,3)\psline[style=latticestyle](3,2)(2,2)
			\psline[style=latticestyle](3,3)(3,4)
			\psline[style=latticestyle](3,4)(2,4)
			\psline[style=latticestyle](4,1)(4,0)\psline[style=latticestyle](4,5)(4,4)
			\psline[style=latticestyle]{<->}(4,4)(3,4)\psline[style=latticestyle](4,4)(5,4)\psline[style=latticestyle](0,4)(1,4)

			\psline[style=polymerstyle](1,3)(1,2)(0.4,2)\psline[style=polymerstyle](4.6,2)(4,2)(3,2)(3,1)
			\psline[style=polymerstyle](4,3)(3,3)(3,4)(4,4)(4,5)\psline[style=polymerstyle](4,0)(4,1)

			\rput(2.5,3){\psline(-0.15,-0.15)(0.15,0.15)\psline(-0.15,0.15)(0.15,-0.15)}
			\psline[style=growthstyle](2,3)(1.7,3.5)(2,4)
		\end{pspicture}
	}
	\\
	\subfloat[][$L_{\mathrm{max}}=3$]{\label{subfig:d}
		\begin{pspicture}(5,5)
			\psline[linewidth=0.5pt](0,0)(5,0)(5,5)(0,5)(0,0)

			\psdot[linewidth=2pt](2,3)
			\psline[style=latticestyle](1,1)(1,0)\psline[style=latticestyle]{<->}(1,1)(0,1)\psline[style=latticestyle]{<->}(5,1)(4,1)\psline[style=latticestyle](1,5)(1,4)
			\psline[style=latticestyle]{<->}(1,2)(1,3)\psline[style=latticestyle]{<->}(1,2)(2,2)
			\psline[style=latticestyle]{<->}(1,4)(1,3)
			\psline[style=latticestyle](2,1)(2,2)\psline[style=latticestyle](2,1)(1,1)
			\psline[style=latticestyle](2,2)(2,3)
			\psline[style=latticestyle](2,3)(2,4)\psline[style=latticestyle](2,3)(3,3)
			\psline[style=latticestyle](2,4)(2,5)\psline[style=latticestyle]{<->}(2,4)(1,4)\psline[style=latticestyle](2,0)(2,1)
			\psline[style=latticestyle]{<->}(3,2)(3,3)\psline[style=latticestyle](3,2)(2,2)
			\psline[style=latticestyle](3,3)(3,4)
			\psline[style=latticestyle](3,4)(2,4)
			\psline[style=latticestyle](4,1)(4,0)\psline[style=latticestyle](4,5)(4,4)
			\psline[style=latticestyle]{<->}(4,4)(3,4)\psline[style=latticestyle](4,4)(5,4)\psline[style=latticestyle](0,4)(1,4)

			\psline[style=polymerstyle](1,3)(1,2)(0.4,2)\psline[style=polymerstyle](4.6,2)(4,2)(3,2)(3,1)
			\psline[style=polymerstyle](4,3)(3,3)(3,4)(4,4)(4,5)\psline[style=polymerstyle](4,0)(4,1)

			\rput(2.5,3){\psline(-0.15,-0.15)(0.15,0.15)\psline(-0.15,0.15)(0.15,-0.15)}
			\psline[style=growthstyle](2,3)(1.7,3.5)(2,4)
			\psline[style=growthstyle](2,4)(1.5,4.4)(1,4)
		\end{pspicture}
	}
	\subfloat[][$L_{\mathrm{max}}=4$]{\label{subfig:e}
		\begin{pspicture}(5,5)
			\psline[linewidth=0.5pt](0,0)(5,0)(5,5)(0,5)(0,0)

			\psdot[linewidth=2pt](2,3)
			\psline[style=latticestyle](1,1)(1,0)\psline[style=latticestyle]{<->}(1,1)(0,1)\psline[style=latticestyle]{<->}(5,1)(4,1)\psline[style=latticestyle](1,5)(1,4)
			\psline[style=latticestyle]{<->}(1,2)(1,3)\psline[style=latticestyle]{<->}(1,2)(2,2)
			\psline[style=latticestyle]{<->}(1,4)(1,3)
			\psline[style=latticestyle](2,1)(2,2)\psline[style=latticestyle](2,1)(1,1)
			\psline[style=latticestyle](2,2)(2,3)
			\psline[style=latticestyle](2,3)(2,4)\psline[style=latticestyle](2,3)(3,3)
			\psline[style=latticestyle](2,4)(2,5)\psline[style=latticestyle]{<->}(2,4)(1,4)\psline[style=latticestyle](2,0)(2,1)
			\psline[style=latticestyle]{<->}(3,2)(3,3)\psline[style=latticestyle](3,2)(2,2)
			\psline[style=latticestyle](3,3)(3,4)
			\psline[style=latticestyle](3,4)(2,4)
			\psline[style=latticestyle](4,1)(4,0)\psline[style=latticestyle](4,5)(4,4)
			\psline[style=latticestyle]{<->}(4,4)(3,4)\psline[style=latticestyle](4,4)(5,4)\psline[style=latticestyle](0,4)(1,4)

			\psline[style=polymerstyle](1,3)(1,2)(0.4,2)\psline[style=polymerstyle](4.6,2)(4,2)(3,2)(3,1)
			\psline[style=polymerstyle](4,3)(3,3)(3,4)(4,4)(4,5)\psline[style=polymerstyle](4,0)(4,1)

			\rput(2.5,3){\psline(-0.15,-0.15)(0.15,0.15)\psline(-0.15,0.15)(0.15,-0.15)}
			\psline[style=growthstyle](2,3)(1.7,3.5)(2,4)
			\psline[style=growthstyle](2,4)(1.5,4.4)(1,4)
			\psline[style=growthstyle](1,4)(0.6,3.5)(1,3)
		\end{pspicture}
	}
	\subfloat[][$L_{\mathrm{max}}=4$]{\label{subfig:f}
		\begin{pspicture*}(5,5)
			\psline[linewidth=0.5pt](0,0)(5,0)(5,5)(0,5)(0,0)

			\psdot[linewidth=2pt](2,3)
			\psline[style=latticestyle](1,1)(1,0)\psline[style=latticestyle]{<->}(1,1)(0,1)\psline[style=latticestyle]{<->}(5,1)(4,1)\psline[style=latticestyle](1,5)(1,4)
			\psline[style=latticestyle]{<->}(1,2)(1,3)\psline[style=latticestyle]{<->}(1,2)(2,2)
			\psline[style=latticestyle]{<->}(1,4)(1,3)
			\psline[style=latticestyle](2,1)(2,2)\psline[style=latticestyle](2,1)(1,1)
			\psline[style=latticestyle](2,2)(2,3)
			\psline[style=latticestyle](2,3)(2,4)\psline[style=latticestyle](2,3)(3,3)
			\psline[style=latticestyle](2,4)(2,5)\psline[style=latticestyle]{<->}(2,4)(1,4)\psline[style=latticestyle](2,0)(2,1)
			\psline[style=latticestyle]{<->}(3,2)(3,3)\psline[style=latticestyle](3,2)(2,2)
			\psline[style=latticestyle](3,3)(3,4)
			\psline[style=latticestyle](3,4)(2,4)
			\psline[style=latticestyle](4,1)(4,0)\psline[style=latticestyle](4,5)(4,4)
			\psline[style=latticestyle]{<->}(4,4)(3,4)\psline[style=latticestyle](4,4)(5,4)\psline[style=latticestyle](0,4)(1,4)

			\psline[style=polymerstyle](1,3)(1,2)(0.4,2)\psline[style=polymerstyle](4.6,2)(4,2)(3,2)(3,1)
			\psline[style=polymerstyle](4,3)(3,3)(3,4)(4,4)(4,5)\psline[style=polymerstyle](4,0)(4,1)

			\rput(2.5,3){\psline(-0.15,-0.15)(0.15,0.15)\psline(-0.15,0.15)(0.15,-0.15)}
			\rput(1,3.5){\psline(-0.15,-0.15)(0.15,0.15)\psline(-0.15,0.15)(0.15,-0.15)}
			\psline[style=growthstyle](2,3)(1.7,3.5)(2,4)
			\psline[style=growthstyle](2,4)(1.5,4.4)(1,4)\psline[style=growthstyle](1,4)(1.5,3.6)(2,4)
		\end{pspicture*}
	}
	\\
	\subfloat[][$L_{\mathrm{max}}=4$]{\label{subfig:g}
		\begin{pspicture*}(5,5)
			\psline[linewidth=0.5pt](0,0)(5,0)(5,5)(0,5)(0,0)

			\psdot[linewidth=2pt](2,3)
			\psline[style=latticestyle](1,1)(1,0)\psline[style=latticestyle]{<->}(1,1)(0,1)\psline[style=latticestyle]{<->}(5,1)(4,1)\psline[style=latticestyle](1,5)(1,4)
			\psline[style=latticestyle]{<->}(1,2)(1,3)\psline[style=latticestyle]{<->}(1,2)(2,2)
			\psline[style=latticestyle]{<->}(1,4)(1,3)
			\psline[style=latticestyle](2,1)(2,2)\psline[style=latticestyle](2,1)(1,1)
			\psline[style=latticestyle](2,2)(2,3)
			\psline[style=latticestyle](2,3)(2,4)\psline[style=latticestyle](2,3)(3,3)
			\psline[style=latticestyle](2,4)(2,5)\psline[style=latticestyle]{<->}(2,4)(1,4)\psline[style=latticestyle](2,0)(2,1)
			\psline[style=latticestyle]{<->}(3,2)(3,3)\psline[style=latticestyle](3,2)(2,2)
			\psline[style=latticestyle](3,3)(3,4)
			\psline[style=latticestyle](3,4)(2,4)
			\psline[style=latticestyle](4,1)(4,0)\psline[style=latticestyle](4,5)(4,4)
			\psline[style=latticestyle]{<->}(4,4)(3,4)\psline[style=latticestyle](4,4)(5,4)\psline[style=latticestyle](0,4)(1,4)

			\psline[style=polymerstyle](1,3)(1,2)(0.4,2)\psline[style=polymerstyle](4.6,2)(4,2)(3,2)(3,1)
			\psline[style=polymerstyle](4,3)(3,3)(3,4)(4,4)(4,5)\psline[style=polymerstyle](4,0)(4,1)

			\rput(2.5,3){\psline(-0.15,-0.15)(0.15,0.15)\psline(-0.15,0.15)(0.15,-0.15)}
			\rput(1.5,4){\psline(-0.15,-0.15)(0.15,0.15)\psline(-0.15,0.15)(0.15,-0.15)}
			\psline[style=growthstyle](2,3)(1.7,3.5)(2,4)
			\psline[linearc=1](2,4)(1.5,4.7)(2,6)\psline[linearc=1](2,-1)(1.5,0.3)(2,1)
		\end{pspicture*}
	}
	\subfloat[][$L_{\mathrm{max}}=4$]{\label{subfig:h}
		\begin{pspicture*}(5,5)
			\psline[linewidth=0.5pt](0,0)(5,0)(5,5)(0,5)(0,0)

			\psdot[linewidth=2pt](2,3)
			\psline[style=latticestyle](1,1)(1,0)\psline[style=latticestyle]{<->}(1,1)(0,1)\psline[style=latticestyle]{<->}(5,1)(4,1)\psline[style=latticestyle](1,5)(1,4)
			\psline[style=latticestyle]{<->}(1,2)(1,3)\psline[style=latticestyle]{<->}(1,2)(2,2)
			\psline[style=latticestyle]{<->}(1,4)(1,3)
			\psline[style=latticestyle](2,1)(2,2)\psline[style=latticestyle](2,1)(1,1)
			\psline[style=latticestyle](2,2)(2,3)
			\psline[style=latticestyle](2,3)(2,4)\psline[style=latticestyle](2,3)(3,3)
			\psline[style=latticestyle](2,4)(2,5)\psline[style=latticestyle]{<->}(2,4)(1,4)\psline[style=latticestyle](2,0)(2,1)
			\psline[style=latticestyle]{<->}(3,2)(3,3)\psline[style=latticestyle](3,2)(2,2)
			\psline[style=latticestyle](3,3)(3,4)
			\psline[style=latticestyle](3,4)(2,4)
			\psline[style=latticestyle](4,1)(4,0)\psline[style=latticestyle](4,5)(4,4)
			\psline[style=latticestyle]{<->}(4,4)(3,4)\psline[style=latticestyle](4,4)(5,4)\psline[style=latticestyle](0,4)(1,4)

			\psline[style=polymerstyle](1,3)(1,2)(0.4,2)\psline[style=polymerstyle](4.6,2)(4,2)(3,2)(3,1)
			\psline[style=polymerstyle](4,3)(3,3)(3,4)(4,4)(4,5)\psline[style=polymerstyle](4,0)(4,1)

			\rput(2.5,3){\psline(-0.15,-0.15)(0.15,0.15)\psline(-0.15,0.15)(0.15,-0.15)}
			\rput(1.5,4){\psline(-0.15,-0.15)(0.15,0.15)\psline(-0.15,0.15)(0.15,-0.15)}
			\psline[style=growthstyle](2,3)(1.7,3.5)(2,4)
			\psline[linearc=1](2,4)(1.5,4.7)(2,6)\psline[linearc=1](2,-1)(1.5,0.3)(2,1)
			\psline[style=growthstyle](2,1)(1.5,1.3)(1,1)
		\end{pspicture*}
	}
	\subfloat[][$L_{\mathrm{max}}=5$]{\label{subfig:i}
		\begin{pspicture*}(5,5)
			\psline[linewidth=0.5pt](0,0)(5,0)(5,5)(0,5)(0,0)

			\psdot[linewidth=2pt](2,3)
			\psline[style=latticestyle](1,1)(1,0)\psline[style=latticestyle]{<->}(1,1)(0,1)\psline[style=latticestyle]{<->}(5,1)(4,1)\psline[style=latticestyle](1,5)(1,4)
			\psline[style=latticestyle]{<->}(1,2)(1,3)\psline[style=latticestyle]{<->}(1,2)(2,2)
			\psline[style=latticestyle]{<->}(1,4)(1,3)
			\psline[style=latticestyle](2,1)(2,2)\psline[style=latticestyle](2,1)(1,1)
			\psline[style=latticestyle](2,2)(2,3)
			\psline[style=latticestyle](2,3)(2,4)\psline[style=latticestyle](2,3)(3,3)
			\psline[style=latticestyle](2,4)(2,5)\psline[style=latticestyle]{<->}(2,4)(1,4)\psline[style=latticestyle](2,0)(2,1)
			\psline[style=latticestyle]{<->}(3,2)(3,3)\psline[style=latticestyle](3,2)(2,2)
			\psline[style=latticestyle](3,3)(3,4)
			\psline[style=latticestyle](3,4)(2,4)
			\psline[style=latticestyle](4,1)(4,0)\psline[style=latticestyle](4,5)(4,4)
			\psline[style=latticestyle]{<->}(4,4)(3,4)\psline[style=latticestyle](4,4)(5,4)\psline[style=latticestyle](0,4)(1,4)

			\psline[style=polymerstyle](1,3)(1,2)(0.4,2)\psline[style=polymerstyle](4.6,2)(4,2)(3,2)(3,1)
			\psline[style=polymerstyle](4,3)(3,3)(3,4)(4,4)(4,5)\psline[style=polymerstyle](4,0)(4,1)

			\rput(2.5,3){\psline(-0.15,-0.15)(0.15,0.15)\psline(-0.15,0.15)(0.15,-0.15)}
			\rput(1.5,4){\psline(-0.15,-0.15)(0.15,0.15)\psline(-0.15,0.15)(0.15,-0.15)}
			\psline[style=growthstyle](2,3)(1.7,3.5)(2,4)
			\psline[linearc=1](2,4)(1.5,4.7)(2,6)\psline[linearc=1](2,-1)(1.5,0.3)(2,1)
			\psline[style=growthstyle](2,1)(1.5,1.3)(1,1)
			\psline[linearc=1](1,1)(0.5,0.3)(1,-1)\psline[linearc=1](1,6)(0.5,4.7)(1,4)
		\end{pspicture*}
	}
	\caption{Example of the growth of a polymer with parameters $d=2$, $N=3$, $L=5$, $k=2$ and $\ell=2$. The $2$ polymers in the system are represented by the bold edges, the dashed arrows represent the underlying graph. \protect\subref{subfig:a}: we start from the root. \protect\subref{subfig:b}: we try the neighbor to the right, and we update the counter $L_{\mathrm{max}}$ to 2. This vertex is occupied by another polymer, so we recoil to the root. Since $\ell > 0$, this move is allowed. \protect\subref{subfig:c}: we try another direction, and go up, which is the only direction left possible since the cross symbolizes the set $D_1$ of directions already tried. \protect\subref{subfig:d}: we try the left vertex, and update $L_{\mathrm{max}}$. \protect\subref{subfig:e}: we try the down vertex, and update $L_{\mathrm{max}}$. This vertex is occupied by another polymer, so we recoil. \protect\subref{subfig:f}: we try the right vertex, which is occupied by the current polymer. Since this is the second try and $k=2$, we have to recoil again. Since $\ell > 1$, this is admissible. If $\ell=1$, this step would yield a failure, because $\max(1, L_{\mathrm{max}}-1) = 3$, and we want to recoil from $v_3$. \protect\subref{subfig:g}-\protect\subref{subfig:i}: we try admissible neighbors that lead to a full polymer.}\label{fig:example_polymer}
\end{figure}

Figure~\ref{fig:example_polymer} on page~\pageref{fig:example_polymer} shows the successful generation of a polymer in dimension $d=2$, with parameters $N=3$, $L=5$, $k=2$ and $\ell=2$.
\\

With this description, it is clear that if the \R procedure returns a polymer $C$, then $C$ and $G$ are compatible. Hence the following property holds:
\begin{proposition}
  If $C$ and $G$ are not compatible, then $P_\RG(C|G) = 0$.
\end{proposition}

We still have to compute the probability $P_\RG(C|G)$ when $C$ and $G$ are compatible. To get an hindsight of what the right answer is, let us consider the case $\ell = 0$: write $C = (v_1, \ldots, v_L)$, and consider any $1 \leq i \leq L-1$. In the process of growing $C$, we have necessarily grown at some point the partial polymer $(v_1, \ldots, v_i)$, and from $v_i$, we have $k$ choices. But since $\ell = 0$, any choice different than $v_{i+1}$ will not lead to $C$. Indeed, if we choose $v' \neq v_{i+1}$, since $\ell = 0$, we will never recoil below $v'$, therefore we will not grow $C$. In opposition, if we choose $v_{i+1}$ for each $1 \leq i \leq L-1$, we will grow $C$: this happens with probability $(1/k)^{L-1}$, which is the answer for $\ell = 0$.
\\

In the case $\ell > 0$, we have still necessarily grown at some point the partial polymer $(v_1, \ldots, v_i)$, and from $v_i$, there are still $k$ potential choices. But in this case, if we choose a vertex $v \neq v_{i+1}$, this is not necessarily the end, because it may happen that we recoil back from $v$ to $v_i$. In this case, and when $k > 1$, we then have a second opportunity to pick the right vertex $v_{i+1}$.

Since $\ell > 0$, when we grow from $v_{i}$ to $v$, $v$ is in the feeler part, and we will eventually recoil back to $v_i$ if and only if $v$ never belongs to the fixed part. Since by definition, $v$ belongs to the fixed part if and only if at some point, we grow the chain $\ell$ steps further, we get that we will recoil back to $v_i$ if and only if the partial polymer $(v_1, \ldots, v_i, v)$ cannot be extended $\ell$ steps further on $G$. Note that this equivalence is true only for vertices that potentially may belong to the fixed part, i.e., for vertices $v_i$ with $i \leq L-\ell$.

For $i \geq L-\ell+1$, this equivalence becomes: we will recoil back to $v_i$ if and only if the partial polymer $(v_1, \ldots, v_i, v)$ cannot be completed.

\begin{definition}
	For $1 \leq i \leq L-1$, a neighbor $v$ of $v_i$ is called admissible if either $i < L-\ell$ and the partial polymer $(v_1, \ldots, v_i, v)$ can be extended $\ell$ steps further, or if $i \geq L-\ell$ and $(v_1, \ldots, v_i, v)$ can be completed to an entire polymer.
	
	\noindent The elementary weight $w_i$ of $v_i$ is the number of admissible neighbors of $v_i$.
\end{definition}

\noindent\textbf{Remark:} In this definition, when the partial polymer $(v_1, \ldots, v_i, v)$ is extended, possible interactions with the remaining part $(v_{i+1}, \ldots, v_L)$ of the polymer are not taken into account. Moreover, note that when $C$ and $G$ are compatible, $v_{i+1}$ is always admissible, because we know that $(v_1, \ldots, v_L) = C$ is possible. Hence in this case, we always have $w_i \geq 1$. And now we can compute the probability $P_\RG(C|G)$:

\begin{definition}
	$W(C|G) = \prod_{1}^{L-1} w_i$ is the \textnormal{weight} of the polymer $C$ given the underlying graph $G$.
\end{definition}

\begin{proposition}
  Given the underlying graph $G$ compatible with $C$, we have
  \begin{align}\label{eq:P_\R}
    P_\RG(C|G) = \frac{1}{W(C|G)} = \left(\prod_{i=1}^{L-1} w_i\right)^{-1}.
  \end{align}
\end{proposition}

\begin{proof}
Let $i \in \{1,\ldots,L-1\}$. In the process of growing $C$, the partial polymer was at some point $(v_1, \ldots, v_i)$. By definition, if we choose as next vertex $v$, then we will eventually recoil back to $v_i$ if and only if $v$ is not admissible. Hence the probability from $v_i$ to make the right choice is one over the number of admissible neighbors, i.e., $1/w_i$.
\end{proof}

\noindent\textbf{Remark:} To compute the elementary weight $w_i$, we have no choice but to use the exhaustive algorithm presented previously. Indeed, we need to know whether under certain constraints, we can grow a self-avoiding path of length $\ell$. So if $\ell$ is large, since we have to do this at most $(k-1)(L-1)$ (when for each $1 \leq i \leq L-1$, we must try to grow a feeler from each of its $k-1$ neighbors) computing the weight can be expensive. However, this is only to be done when we have successfully grown a polymer. The complexity here is linear in $k$ and $L$, but exponential in $\ell$.

\section{Properties of the Recoil Growth Algorithm}\label{sec:properties}

\begin{definition}
	The density $\rho = NL/a^d$ of the system is the ratio of the number of vertices occupied over the total number of vertices.
\end{definition}

In this section, we prove the following theorem, where it is assumed that the size $a$ of the underlying lattice satisfies $a \geq L$, which is the relevant case in practice:

\begin{theorem}
With the right probability of acceptance $P_\acc\big((S_\o, G_\o), (S_\n, G_\n)\big)$, the \R algorithm is reversible with stationary distribution $q$. The algorithm is moreover irreducible at densities smaller than $\lfloor a/L\rfloor/a$.
\end{theorem}


To prove this theorem, we first deal with the stationary distribution in the next section, and then discuss some issues related to the irreducibility. Note that we prove the existence of a non-empty region for which both the original RG and the \R algorithms are irreducible.

\subsection{Reversibility}

\begin{lemma}\label{lemma:balance_equations}
  Take any two states $S_\o$ and $S_\n$ in $\mathcal S$ that differ by only one polymer $C_\o$ in $S_\o$ and $C_\n$ in $S_\n$, and take any two underlying graphs $G_\o$ and $G_\n$ compatible with $C_\o$ and $C_\n$ respectively. If we choose
  \begin{align}\label{eq:proba-acc}
    P_\acc\big((S_\o, G_\o), (S_\n, G_\n)\big) = \min\left(1, \frac{q(S_\n) W(C_\n | G_\n)}{q(S_\o) W(C_\o | G_\o)}\right),
  \end{align}
  then $q$ satisfies the detailed balance equations:
  \begin{align*}
  	q(S_\o) P(S_\o \rightarrow S_\n) = q(S_\n) P(S_\n \rightarrow S_\o).
  \end{align*}
\end{lemma}

\begin{proof}
Recall that we have from~\eqref{eq:transition}
\begin{align*}
  P(S_\o \rightarrow S_\n) = \frac{1}{N}\sum_{G_\n,\ G_\o}P_\U(G_\n) P_\RG(C_\n|G_\n)P_\C(G_\o|C_\o) P_\acc\big((S_\o, G_\o), (S_\n, G_\n)\big).
\end{align*}
We can show that $q$ not only satisfies the detailed balance equations, but satisfies a stronger ``local'' condition: if we define
\begin{align*}
  P\big((S_\o, G_\o) \rightarrow (S_\n, G_\n)\big) = P_\U(G_\n) P_\RG(C_\n|G_\n)P_\C(G_\o|C_\o) P_\acc\big((S_\o, G_\o), (S_\n, G_\n)\big),
\end{align*}
then $P(S_\o \rightarrow S_\n) = 1/N\sum_{G_\n,\ G_\o} P\big((S_\o, G_\o) \rightarrow (S_\n, G_\n)\big)$, and it is clearly sufficient to show that for all $S_\o, S_\n, G_\o$ and $G_\n$, $q$ satisfies
\begin{align}\label{eq:micro-rev}
  q(S_\o) P\big((S_\o, G_\o) \rightarrow (S_\n, G_\n)\big) = q(S_\n) P\big((S_\n, G_\n) \rightarrow (S_\o, G_\o)\big).
\end{align}
But we have seen that $P_\U(G_\n) P_\C(G_\o|C_\o) = P_\U(G_\o) P_\C(G_\n|C_\n)$ because there exists a constant $\eta$ such that $P_\C(G|C) = \eta P_\U(G)$, so~\eqref{eq:micro-rev} amounts to
\begin{align*}
	\frac{P_\acc\big((S_\o, G_\o), (S_\n, G_\n)\big)}{P_\acc\big((S_\n, G_\n), (S_\o, G_\o)\big)} = \frac{q(S_\n)W(C_\n|G_\n)}{q(S_\o)W(C_\o|G_\o)},
\end{align*}
what is true by choice of the probability of acceptance.
\end{proof}

\subsection{On the irreducibility of the Markov Chain}\label{sub:irreducibility}

Knowing that $q$ satisfies the detailed balance equations is not enough for the algorithm to work properly: one needs results on the irreducibility of the Markov Chain as well. This question seems to be a hard problem, and we aim at giving some hints in this section.

\subsubsection{Proof of theorem 1}

\begin{lemma}
  For any $d, N, L$ and $a \geq L$, the chain is irreducible if $\rho \leq \lfloor a/L\rfloor/a $.
\end{lemma}

\begin{proof}
The idea is to show that from any configuration $S \in {\mathcal S}$, we can reach a particular configuration $\widetilde S$ where all the polymers are straight, and lie in some specific boxes. The boxes are defined as follows.

$\G = (\Z / a\Z)^d$ is a $d$-dimensional cube, and for any $x \in (\Z / a\Z)^{d-1}$, we consider the subset $F_x = \{y = (y_i) \in \G: (y_1, \ldots, y_{d-1}) = x, y_d \in a\Z\}$ where the $d-1$ first coordinates are fixed. $F_x$ is in bijection with $\Z / a \Z$. Since $a \geq L$, we write $a = nL + r$ with $n = \lfloor a/L \rfloor \geq 1$ and $0 \leq r \leq L-1$. For each $x$, we can then divide $F_x$ into $n$ boxes, the first box being in bijection with $\{0, \ldots, L-1\}$, the second with $\{L, \ldots, 2L-1\}$, etc\ldots The total number of boxes $B$ is equal to $a^{d-1}n$, therefore we get, using the definition of $\rho$ and the hypotheses:
\begin{align*}
	\frac{B}{NL} = \frac{n}{\rho a} \geq 1.
\end{align*}
But $NL$ is the number of occupied vertices, so the previous inequality means that we have more boxes than occupied vertices. We study only the worst case, where $B=NL$: one of two things may happen.

If each box is occupied by exactly one vertex, then each box is intersected by exactly one polymer. So we can consider each polymer in turn, and put each one in some box that it occupies. Each polymer will then be straight and lie in some box.

If some boxes are occupied by more than one vertex, then necessarily, some boxes are empty. Then we can consider the polymers in turn and assign them to empty boxes while we have empty boxes. We stop either when we have no more empty boxes or when every polymer has been assigned to a box. In the latter case, we are done. And it is easy to see that the former case never happens, because by putting a polymer in a box, we create empty boxes.

In any case, we can go to a state where all the polymers are in boxes. Such states are clearly connected, and the chain is therefore irreducible.
\end{proof}

\noindent\textbf{Remark:} With regards to simulation, the longer the polymers, the better. In~\cite{Consta}, the authors investigate cases when $L = 100$. So this result insures irreducibility at very low densities, around $0.01$, compared to the values we would be interested in, like $0.6$. Moreover, one question that is still open at this point is to give a lower bound independent of the parameters of the system. For instance, it seems a reasonable bet that the algorithm is always irreducible for $\rho = 10^{-10}$.

\subsubsection{Informal discussion}

In this section, we give examples of parameters for which the algorithm is irreducible, some for which it is not, and we state some conjectures.
\\

First, observe that if two states $S$ and $S'$ differ by only one polymer, then the probability $P(S \rightarrow S')$ of going from $S$ to $S'$ in one step of the algorithm is strictly positive for any  $k$ and $\ell$. These two parameters obviously influence the value of such a probability, but it will nevertheless always be positive. The question of irreducibility does therefore not depend on these two parameters.

For purposes of irreducibility, the algorithm is then characterized by $N, L$ and $\G$. Recall that $\G = \left(\Z / a \Z\right)^d$ is a cyclic cube in dimension $d$. Since $\rho = NL/a^d$, we see that the \R algorithm is characterized by five parameters $N, L, \rho, a$ and $d$, so we adopt the notation $RG^*_d(N,L,a,\rho)$ to refer to the \R algorithm with these parameters. Any one of these parameters is uniquely determined by the four others thanks to the relation $\rho = NL/a^d$.

Before going on, we rule out trivial cases that are of no interest: for $N = 1$ or $L=1$, the algorithm is always irreducible. So in the rest of this discussion, we always assume $N, L \geq 2$. The case $L=2$ corresponds to the so-called ``monomer-dimer system'', which has received considerable attention in the literature, e.g.~\cite{dimer-phase-transition, dimer-sampling}, with~\cite{dimer-phase-transition} dealing with problems close to the ones we are interested in. However, the Metropolis algorithm proposed in~\cite{dimer-phase-transition} allows to remove or add dimers, in which case the problem of irreducibility is trivial.

Proposition~7 deals with a simple case:
\begin{proposition}
	If $\rho = 1$, the algorithm is never irreducible.
\end{proposition}

\begin{proof}
Assume that $\rho = 1$ and you start from some state $S$. Then removing a polymer $C$ leaves exactly $L$ free vertices. Hence for any state $S'$ where $C$ has been replaced by $C'$, $C$ and $C'$ necessarily occupy the same set of vertices. In particular, two vertices $v$ and $v'$ neighbors in $\G$ which belong to different polymers in $S$ will never be part of the same polymer. Since there exist such states, $RG^*_d(N, L,a,1)$ is not irreducible.
\end{proof}

The next two propositions will help us build our intuition:

\begin{proposition}
	For any $\rho < 1$ and any other parameters, $RG^*_d(N,2,a,\rho)$ is irreducible.
\end{proposition}

\begin{proof}
We deal with the case $d=2$, and when there is only one free vertex. When there are more free vertices or when $d > 2$, the same ideas can be extended to prove that the system is still irreducible. So from now on, we assume $d=2$ and that there is one free vertex.

We show that from any state, we can reach the state depicted in Figure~\ref{fig:dimer}, where all the dimers are horizontal, except on one column where they all are vertical. The idea is the following: we consider the number $\kappa$ of dimers that are ``vertical'', i.e., the number of dimers that are aligned along a column and not along a line. We will show that there exists a path, i.e., a sequence of moves, along which $\kappa$ is decreasing. Figure~\ref{fig:dimer} represents a state where $\kappa$ is equal to its minimal value.

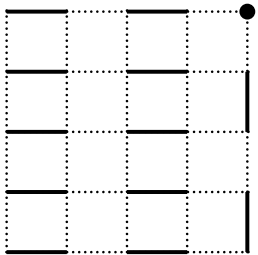
\begin{figure}
\begin{center}
	\psset{unit=0.8,linewidth=1.5pt,angle=90}
	\begin{pspicture}(4,4)
		\psgrid[griddots=10,gridwidth=1pt,subgriddiv=1,gridlabels=0](0,0)(4,4)
		\rput(0,0){\psline(0,0)(1,0)}\rput(2,0){\psline(0,0)(1,0)}
		\rput(0,1){\psline(0,0)(1,0)}\rput(2,1){\psline(0,0)(1,0)}
		\rput(0,2){\psline(0,0)(1,0)}\rput(2,2){\psline(0,0)(1,0)}
		\rput(0,3){\psline(0,0)(1,0)}\rput(2,3){\psline(0,0)(1,0)}
		\rput(0,4){\psline(0,0)(1,0)}\rput(2,4){\psline(0,0)(1,0)}
		\psline(4,0)(4,1)\psline(4,2)(4,3)
		\psdot[linewidth=2pt](4,4)
	\end{pspicture}
\end{center}	
	\vspace{-0.5cm}
	\caption{State where all the dimers are horizontal, except for one column, which contains the ``whole'', i.e., the free vertex.}\label{fig:dimer}
\end{figure}

For the parameters considered, $a^2 = 2N+1$, hence $a$ is odd. Since only one vertex is free, there is only one line on which there is a free vertex. So for the $a-1$ other lines, the $a$ vertices are occupied by dimers. Since $a$ is odd, and since a horizontal dimer occupies an even number of vertices, there is an odd number of vertical dimers that have exactly one vertex lying on this line. In particular, there is at least one vertical dimer on each of these $a-1$ lines.
\\

\begin{figure}
	\psset{linewidth=1.5pt}
	\vspace{0.4cm}
	\subfloat[][Move \texttt{a}]{\label{subfig:el_move_1}
		\begin{pspicture}(4,2)
			\psgrid[griddots=10,gridwidth=1pt,subgriddiv=1,gridlabels=0](0,0)(4,0)\rput(0,0){\psline(0,0)(1,0)}\rput(2,0){\psline(0,0)(1,0)}\psdot[linewidth=2pt](4,0)
			\psline{->}(2,1.5)(2,0.5)
			\psgrid[griddots=10,gridwidth=1pt,subgriddiv=1,gridlabels=0](0,2)(4,2)\rput(0,2){\psline(0,0)(1,0)}\psdot[linewidth=2pt](2,2)\rput(3,2){\psline(0,0)(1,0)}
		\end{pspicture}
	}
	\hspace{0.5cm}
	\subfloat[][Move \texttt{b}]{\label{subfig:el_move_2}
		\begin{pspicture}(2,2)
			\psgrid[griddots=10,gridwidth=1pt,subgriddiv=1,gridlabels=0](0,0)(0,2)\psdot[linewidth=2pt](0,0)\psline(0,1)(0,2)
			\psline{->}(0.5,1)(1.5,1)
			\psgrid[griddots=10,gridwidth=1pt,subgriddiv=1,gridlabels=0](2,0)(2,2)\psdot[linewidth=2pt](2,2)\psline(2,0)(2,1)
		\end{pspicture}
	}
	\hspace{0.5cm}
	\subfloat[][Move \texttt{c}]{\label{subfig:el_move_3}
		\begin{pspicture}(4,2)
			\psgrid[griddots=10,gridwidth=1pt,subgriddiv=1,gridlabels=0](0,0)(1,1)\psdot[linewidth=2pt](0,0)\psline(1,0)(1,1)
			\psline{->}(1.5,0.5)(2.5,0.5)
			\psgrid[griddots=10,gridwidth=1pt,subgriddiv=1,gridlabels=0](3,0)(4,1)\psdot[linewidth=2pt](4,1)\psline(3,0)(4,0)
		\end{pspicture}
	}
	\caption{The three elementary moves: \texttt{a} and \texttt{b} do not change $\kappa$, whereas \texttt{c} decreases $\kappa$ by one.}\label{fig:elementary_moves}
\end{figure}
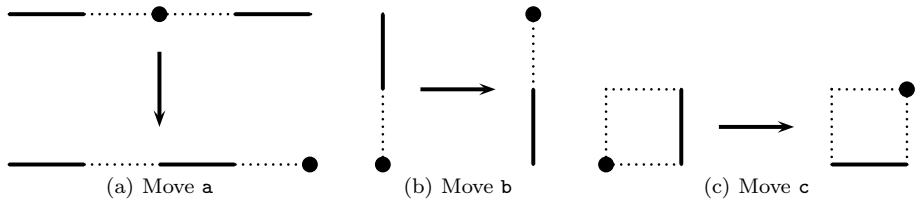

We now describe the path along which $\kappa$ is decreasing. This path is made of the three elementary moves \texttt{a}, \texttt{b} and \texttt{c} depicted in Figure~\ref{fig:elementary_moves}. We consider the line with the whole: there is an even number of vertices occupied, therefore there is an even number of vertical dimers sharing a vertex with this line, which is possibly null.

If there are at least two such dimers, we can move to the right thanks to move~\texttt{a} until the vertex to the right of the whole is a vertical dimer, as in case~\subref{subfig:el_move_3} of Figure~\ref{fig:elementary_moves}. Then by applying move~\texttt{c}, we can reduce $\kappa$ by one. The whole is then on a new line.

If there is no vertical dimer on the line of the whole, we can change line: first, we apply move~\texttt{a} to go to the right until the whole is under some vertical dimer as in case~\subref{subfig:el_move_2}. There exists such a dimer, because there is an odd number of vertical dimers on the above line. Then we apply move~\texttt{b}, and we jumped two lines. Since there is an odd number of lines, by jumping two lines at a time, we can put the whole on any line (this is the same reason why you can always go under a vertical dimer).
\\

So now, we know that from any state, we can reach a state where $\kappa$ is minimum. It is not hard to see that such a state is as follows: in any line except the line with the whole, there is exactly one vertical dimer. To reach the state of Figure~\ref{fig:dimer}, one just has to align these $(a-1)/2$ vertical dimers: it is not hard to do so by combining moves~\texttt{a} and~\texttt{c}, so we are done.
\end{proof}

\begin{figure}
\begin{center}
	\psset{unit=1,linewidth=1.5pt}
	\begin{pspicture*}(6,6)
		\psgrid[griddots=10,gridwidth=1pt,subgriddiv=1,gridlabels=0](0,0)(6,6)
		\psline(0,3)(2,3)(2,5)
		\psline(3,6)(3,4)(5,4)
		\psline(6,3)(4,3)(4,1)
		\psline(3,0)(3,2)(1,2)
		\psdot[linewidth=2pt](3,3)
		\psline[fillcolor=red,fillstyle=vlines](-0.5,3.5)(1.5,3.5)(1.5,5.5)(2.5,5.5)(2.5,6.5)(-0.5,6.5)(-0.5,3.5)
		\psline[fillcolor=red,fillstyle=vlines](3.5,4.5)(5.5,4.5)(5.5,3.5)(6,3.5)(6,6.1)(3.5,6.1)(3.5,4.5)
		\psline[fillcolor=red,fillstyle=vlines](-0.5,-0.5)(-0.5,2.5)(0.5,2.5)(0.5,1.5)(2.5,1.5)(2.5,-0.5)(-0.5,-0.5)
		\psline[fillcolor=red,fillstyle=vlines](6.1,-0.1)(6.1,2.5)(4.5,2.5)(4.5,0.5)(3.5,0.5)(3.5,-0.1)(6.1,-0.1)
	\end{pspicture*}
\end{center}	
	\vspace{-0.5cm}
	\caption{Example which shows that $RG^*_2(N, 5,a,80/81)$ is not irreducible: there is no free vertex in the dashed areas, so polymers in the dashed areas cannot move. Nor can the polymers surrounding the only free vertex.}\label{fig:counter-example}
\end{figure}
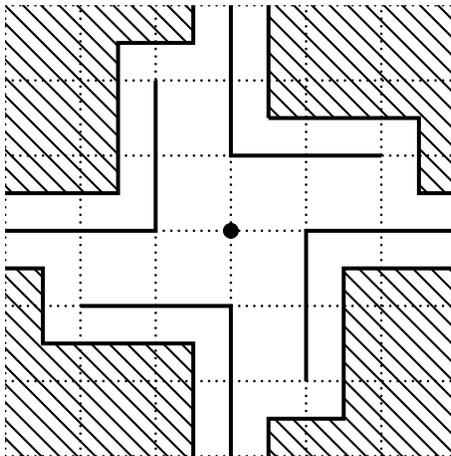

\begin{proposition}
	$RG^*_2(N,5,a,80/81)$ is not irreducible.
\end{proposition}

\begin{proof}
We have $80/81 = 5N/a^2$, which leads by simple considerations to $N=16n^2$ for some integer $n>0$, and $a=9n$. Now we restrict our attention to the case $n=1$, in which case there are $81$ vertices, so only one vertex is free. Figure~\ref{fig:counter-example} zooms on a region of a state of such a system, where the polymers have a specific configuration around the only free vertex. Every vertex in any of the dashed area is occupied. Hence from this state, two things may happen.

If we remove a polymer that is in the dashed area, the only area where we can grow another polymer is in the region freed by this polymer. Hence in this case, every polymer still occupies the same set of vertices.

But if we remove a polymer surrounding the free vertex, then it is easy to see that the only possible way to grow it is at the exact same place. Indeed, simple considerations show that a new polymer cannot occupy this free vertex, because it acts like a dead-end: from this vertex, the polymer would have to pick one of two possible directions, and whatever the direction chosen, there is not enough room to grow the polymer.

Hence from this state, the only states reachable are states where each polymer occupies the same set of vertices, so the system is not irreducible.

For $n > 1$, we can repeat over space the state obtained in the case $n=1$ to obtain a state that has the exact same property, hence the proposition is proved.
\end{proof}

From these two examples, we see that the parameter $\rho$ is not sufficient to characterize the irreducibility of the system, meaning that it can happen that $RG^*_d(N,L,a,\rho)$ is irreducible, whereas $RG^*_d(N',L',a',\rho')$ with $\rho' < \rho$ is not. However, it seems intuitively that the critical parameter is actually the length of the polymers, and so one could wonder whether for fixed length, and for fixed dimension $d$, the density is not enough to characterize the irreducibility. This idea is the object of the following conjecture:

\begin{conjecture}
	If $RG^*_d(N,L,a,\rho)$ is irreducible, then $RG^*_d(N',L,a',\rho')$ is irreducible for any $N', a'$ such that $\rho' < \rho$.
\end{conjecture}
If this conjecture were to be true, then one could consider the quantity
\begin{align*}
	\rho_d(L) = \sup_{N, a}\{\rho: RG^*_d(N,L,a,\rho) \textrm{ is irreducible}\}.
\end{align*}
For any $N$, $a$, $RG^*_d(N,L,a,\rho)$ would be irreducible if and only if $\rho < \rho_d(L)$, and a second conjecture, that would reflect the fact that the longer the polymers, the harder it is for the system to be irreducible, would be:
\begin{conjecture}
	$\rho_d(L)$ is decreasing in $L$.
\end{conjecture}
Finally, we state the the most natural, and probably easiest, conjecture: it says that if the system is irreducible, then removing polymers, shortening the polymers or growing the box should leave the system irreducible.
\begin{conjecture}
	Suppose that $RG^*_d(N,L,a,\rho)$ is irreducible. Then $RG^*_d(N',L',a',\rho')$ is irreducible for any $N' \leq N, L' \leq L$ and $a \geq a'$, where the value of $\rho'$ is determined by the other parameters.
\end{conjecture}

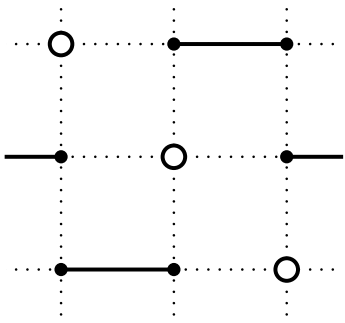
\begin{figure}
	\begin{center}
		\psset{linewidth=1.5pt,unit=1.5}
		\begin{pspicture}(4,4)
			\psgrid[griddots=10,gridwidth=1pt,subgriddiv=1,gridlabels=0](0,0)(4,4)
			\psline[linecolor=white,fillstyle=solid,fillcolor=white](-0.1,-0.1)(-0.1,4.1)(0.5,4.1)(0.5,-0.1)(-0.1,-0.1)
			\psline[linecolor=white,fillstyle=solid,fillcolor=white](-0.1,-0.1)(-0.1,0.5)(4.1,0.5)(4.1,-0.1)(-0.1,-0.1)
			\rput(3.6,0){\psline[linecolor=white,fillstyle=solid,fillcolor=white](-0.1,-0.1)(-0.1,4.1)(0.5,4.1)(0.5,-0.1)(-0.1,-0.1)}
			\rput(0,3.6){\psline[linecolor=white,fillstyle=solid,fillcolor=white](-0.1,-0.1)(-0.1,0.5)(4.1,0.5)(4.1,-0.1)(-0.1,-0.1)}
			\psline(1,1)(2,1)\psdot(1,1)\psdot(2,1)\pscircle[fillstyle=solid,fillcolor=white](1,3){0.1}
			\rput(1,2){\psline(1,1)(2,1)\psdot(1,1)\psdot(2,1)}\pscircle[fillstyle=solid,fillcolor=white](3,1){0.1}
			\pscircle[fillstyle=solid,fillcolor=white](2,2){0.1}\psdot(1,2)\psdot(3,2)\psline(1,2)(0.5,2)\psline(3,2)(3.5,2)
		\end{pspicture}
	\end{center}
	\vspace{-1cm}
	\caption{Example of a state in $RG^*_2(N,L,a,\rho)$ that cannot be extended to $RG^*_2(N+1,L,a,\rho)$, although $RG^*_2(N+1,L,a,\rho)$ is irreducible ($N=3,L=2$ and $a=3$). Circles represent unoccupied vertices.}\label{fig:counter_example_proof}
\end{figure}

Note that this conjecture would immediately imply Conjecture 2. A natural idea to try to prove this last conjecture is the following: suppose for instance that you want to show that $RG^*_d(N-1,L,a,\rho')$ is irreducible, given that $RG^*_d(N,L,a,\rho)$ is. Then you could add a fake polymer, make any move in $RG^*_d(N,L,a,\rho)$, remove the fake polymer, and you are done. However, it is not always possible to add a polymer to $RG^*_d(N-1,L,a,\rho')$, even if $RG^*_d(N,L,a,\rho)$ is irreducible, as shown in the simple example on Figure~\ref{fig:counter_example_proof}. Hence, Conjecture 3 would be true if from any state in $RG^*_d(N-1,L,a,\rho)$, we could reach a state where we can add a fake polymer. Similar ideas hold for $L$ and $a$.

\subsection{Extensions of the Recoil Growth Algorithm}\label{sub:extensions}

\subsubsection{Polymers of different lengths}\label{subsub:length}

All what we have done still holds when the polymers have different lengths, under the reasonable condition that we always try to replace a polymer by another polymer with the same length. Then it is easy to check that all the above formulas still hold, where each time $L$ is to be replaced by the length $L_C$ of the polymer $C$ that we are trying to replace, and the constant $\gamma$ is to be replaced by a number $\gamma_{L_C}$.
\\

When the polymers have different length, a new problem arises which we have not mentioned so far: the polymers are then \textit{de facto} distinguishable. Distinguishability influences the irreducibility of the algorithm. A trivial example is when $L=1$ and $\rho=1$, i.e., the full lattice is covered by its $N$ vertices as $N$ polymers: if the polymers are indistinguishable, the system is irreducible, because there is only one state. But if they are distinguishable, then the system is not irreducible, because there are $N!$ isolated states.

\subsubsection{Underlying graphs with random out-degrees}\label{subsub:random}

In this section, we study an extension proposed in~\cite{Consta} that relaxes the constraint that $k$ needs to be integer-valued, and adopt it in our framework. Although for the original RG algorithm, this does not require to change the probability of acceptance $P_\acc$, for the \R algorithm, it does. We compute below the correct value for this new algorithm, to which we refer as the \emph{extended \R algorithm}.
\\

In this extension, the out-degree of each vertex of an underlying graph is random instead of deterministic, so the definition of an underlying graph given earlier does not apply in this section. However, the generation of underlying graphs, compatible or not, is very similar as in the case where the out-degrees are deterministic. There is only one additional step: instead of assigning $k$ or $k-1$ uniformly at random, we first draw a number $1 \leq \kappa \leq Q$ with probability $p_\kappa$, and then we pick $\kappa$ or $\kappa-1$ vertices uniformly at random. The probability distribution $p$ is therefore an additional parameter to the algorithm. Note that we impose $p_0 = 0$, because in the process of growing a compatible underlying graph, we have to assign at least one out-edge to the vertices on the polymer.

The interest of this extension is to allow any mean random degree $<k> = \sum \kappa p_\kappa$, so that the optimization of the algorithm can be more efficient.

\begin{definition}
	$W^0(C|G)$ is the weight of $C$ on the compatible underlying graph $G$ under $\ell = 0$. In other words, $W^0(C|G) = \prod_{i=1}^{L-1} d(v_i)$ if $C = (v_1, \ldots, v_L)$ and $d(v)$ is the out-degree of $v \in G$.
\end{definition}

We use the same notations as in Lemma~\ref{lemma:balance_equations} to state the result of this section:

\begin{lemma}
	If one sets
	\begin{align}\label{eq:proba-acc-ext}
		P_\acc\big((S_\o, G_\o), (S_\n, G_\n)\big) = \min\left(1, \frac{q(S_\n) W(C_\n|G_\n) W^0(C_\n|G_\n)^{-1}}{q(S_\o) W(C_\o|G_\o) W^0(C_\o|G_\o)^{-1}}\right),
	\end{align}
	then $q$ is the stationary distribution of the extended \R algorithm.
\end{lemma}

\begin{proof}
We show that $q$ satisfies the same ``locale'' balance equations~\eqref{eq:micro-rev}. The probability of acceptance $P_\acc$ is given, so we have to compute the new probabilities of generating our objects under the extended \R algorithm, which correspond to formulas~\eqref{eq:P_U}, \eqref{eq:P_C} and~\eqref{eq:P_\R}.

Since only the generation of the underlying graphs changes in this algorithm, formula~\eqref{eq:P_\R} still holds with the same definition for the $w_i$. Note that $w_i$ is not bounded by $k$ anymore, but rather by the out-degree $d(v_i)$ of $v_i$. More generally, we note $d(v)$ the out-degree of $v$, with no possible confusion on the underlying graph that is considered.
\\

We need some notations: let $C$ be a polymer, $G$ a compatible underlying graph, and $1 \leq i \leq Q$. We note $\alpha_i = \binom{Q}{i}^{-1}$, $\beta_i = \binom{Q-1}{i-1}^{-1}$. $\widetilde C$ is $C$ minus its extremity that is not the root of $G$, $k_i(G) = |\{v \in G: d(v) = i\}|$, $q_i(G|C) = |\{v \in G \backslash \widetilde C: d(v) = i\}|$ and $\tilde q_i(G|C) =|\{v \in \widetilde C: d(v) = i\}|$.
\\

Now if we turn our attention to the generation of a general underlying graph, we see that formula~\eqref{eq:P_U} becomes
\begin{align*}
	P_\U(G) = \frac{1}{\gamma} \prod_{v \in G} p_{d(v)} \alpha_{d(v)} = \frac{1}{\gamma} \prod_{i=1}^Q (p_i \alpha_i)^{k_i(G)}.
\end{align*}
Indeed, we assign the out-degree $d$ to the vertex $v$ with probability $p_d$, and then, there are $1/\alpha_d$ different ways to choose its neighbors. Note that $p_i = 0$ implies $k_i(G) = 0$, and we adopt the convention that $0^0 = 1$. For the same reasons, formula~\eqref{eq:P_C} becomes
\begin{align*}
	P_\C(G|C) = \frac{1}{2} \prod_{v \in G\backslash \widetilde C} p_{d(v)} \alpha_{d(v)} \prod_{v \in \widetilde C} p_{d(v)} \beta_{d(v)} = \frac{1}{2} \prod_{i=1}^Q \left[(p_i \alpha_i)^{q_i(G|C)} (p_i \beta_i)^{\tilde q_i(G|C)}\right].
\end{align*}
Using $|\widetilde C| = L-1$, $k_i(G) = q_i(G|C) + \widetilde q_i(G|C)$, $\beta_i = Q \alpha_i / i$ and $W^0(G|C) = \prod_{v \in \widetilde C} d(v) = \prod_1^{Q} i^{\widetilde q_i(G|C)}$, we finally can rewrite
\begin{align*}
	P_\C(G|C) = \frac{\gamma Q^{L-1}}{2} P_\U(G) W^0(C|G)^{-1}.
\end{align*}
Putting all these formulas together, it is then easy to check that~\eqref{eq:micro-rev} amounts to
\begin{align*}
	\frac{P_\acc\big((S_\o, G_\o), (S_\n, G_\n)\big)}{P_\acc\big((S_\n, G_\n), (S_\o, G_\o)\big)} = \frac{q(S_\n)W(C_\n|G_\n)W^0(C_{\mathrm n}|G_{\mathrm n})^{-1}}{q(S_\o)W(C_\o|G_\o)W^0(C_{\mathrm o}|G_{\mathrm o})^{-1}}.
\end{align*}
The acceptance probability chosen exactly satisfies this equality.
\end{proof}

\noindent\textbf{Remark:} When $p$ is supported by only one point $k$, the terms $W^0$ cancel out, and we find the probability of acceptance for the standard \R algorithm.

\section{Implementation}\label{sec:implementation}

We have mentioned at several occasions throughout this paper that the original algorithm in~\cite{Consta} is actually an efficient implementation of the \R algorithm as defined in the present paper. In this section, we present this implementation, cast in our framework.

The inefficiency of the \R algorithm comes from the generations of the underlying graphs. As we have defined them, the only way they interact with the rest of the system is through their roots. The root of an underlying graph has to be a free vertex, but once it is chosen, we do not care whether vertices which we add belong to existing polymers or not, nor do we care whether they are at a distance larger than $L$ from the root. However, we will never visit such vertices in the process of growing a polymer.

We have already stressed out that the underlying graph and the growth of the polymer are independent concepts. This has made possible to define and study them separately. But since they are independent, nothing forbids to generate the underlying graph while growing the polymer. This way, we are sure to only generate the parts of the underlying graph that we visit during the growth procedure. So the idea is just to grow the polymer as usual, and to complete the underlying graph when the current endpoint of the polymer has not been assigned out-edges yet.

If $d(v)$ is the out-degree of vertex $v$, the entangled \R algorithm becomes as follows:

\noindent\rule{\textwidth}{0.3mm}
\vspace{-0.4cm}
\texttt{
	\begin{enumerate}[leftmargin=*, rightmargin=0.1cm]
		\item Initialization:
		\begin{enumerate}
			\item Pick an unoccupied vertex $r$ uniformly at random
			\item Set $T = \{r\}$ and $V = E = \emptyset$
			\item Set $\overline C = (r), D_1 = \emptyset$ and $L_{\mathrm{max}} = 1$
		\end{enumerate}
		\item At each step, with $\overline C = (v_1, \ldots, v_i)$:
		\begin{enumerate}
			\item If $i = L$, stop, $\overline C$ is a complete polymer
			\item If $v_i$ has not been assigned out-edges, i.e.\!\!\! $v_i \in T$:
			\begin{enumerate}
				\item Set $d(v_i) = \kappa$ with probability $\pi_\kappa$
				\item Choose uniformly at random $d(v_i)$ different vertices among the $Q$ neighbors of $v_i$ in $\G$
				\item For each vertex $v$ chosen:
				\begin{enumerate}
					\item Add the edge $v_i \rightarrow v$ to $E$
					\item If $v \notin T$ and $v \notin E$, add $v$ to $T$
				\end{enumerate}
				\item Remove $v_i$ from $T$ and add it to $V$
			\end{enumerate}
			\item If $|D_i| = d(v_i)$, recoil:
			\begin{enumerate}
				\item If $i > \max(1, L_{\mathrm{max}}-\ell)$, set $\overline C = (v_1, \ldots, v_{i-1})$ and go to 2
				\item Otherwise, stop, the generation has failed
			\end{enumerate}
			\item Else keep picking uniformly at random a vertex $v$ neighbor of $v_i$ in $G$ and not in $D_i$, add it to $D_i$, and stop when either $v$ is unoccupied or $|D_i| = d(v_i)$: 
			\begin{enumerate}
				\item If $v$ is unoccupied
				\begin{enumerate}
					\item Set $\overline C = (v_1, \ldots, v_i, v), D_{i+1} = \emptyset$ and update $L_{\mathrm{max}}$
					\item If $v \notin T$ and $v \notin E$, add $v$ to $T$
					\item Go to 2
				\end{enumerate}
				\item Else, $|D_i| = d(v_i)$, and recoil as specified in 2.(c)
			\end{enumerate}
		\end{enumerate}
	\end{enumerate}
}
\vspace{-0.2cm}
\noindent\rule{\textwidth}{0.3mm}
\\

At the end of this procedure, it can happen that the underlying graph is not complete: some vertices can have not been assigned out-edges. It is then important to keep the current partial underlying graph in memory, because this graph will be further completed if the next step of the algorithm is to compute the weights. Indeed, the same idea applies in this case: when we compute the weights, we do as usual, and we complete the current partial underlying graph when required.

\subsection*{Acknowledgment}

The author would like to thank Persi Diaconis and Philippe Robert for their persistent support and interest in this work, as well as Fr\'{e}d\'{e}ric Meunier for very fruitful discussions.

\providecommand{\bysame}{\leavevmode\hbox to3em{\hrulefill}\thinspace}
\providecommand{\MR}{\relax\ifhmode\unskip\space\fi MR }
\providecommand{\MRhref}[2]{%
  \href{http://www.ams.org/mathscinet-getitem?mr=#1}{#2}
}
\providecommand{\href}[2]{#2}

\end{document}